\documentclass[9pt,
aps,prl,
reprint,
 superscriptaddress,
]{revtex4-1}
\usepackage{dcolumn}
\usepackage{bm}
\usepackage{xr}
\usepackage%
{graphicx}
\usepackage{xfrac}
\usepackage[nice]{nicefrac}
\usepackage{extdash}
\usepackage{epsfig}
\usepackage{isomath}
\usepackage{textcomp}
\usepackage{bm}
\usepackage[english]{babel}
\usepackage{todonotes}
\usepackage{color,soul}
\usepackage{caption}
\usepackage{verbatim}
\usepackage{times}
\usepackage{amsmath}

\DeclareFontFamily{U}{euc}{}
\DeclareFontShape{U}{euc}{m}{n}{<-6>eurm5<6-8>eurm7<8->eurm10}{}%
\DeclareSymbolFont{AMSc}{U}{euc}{m}{n} 
\DeclareMathSymbol{\umu}{\mathord}{AMSc}{"16} 
\newcommand{\ensuretext}[1]{\ensuremath{\text{#1}}}

\newcommand{\unit}[1]{\ensuretext{\textrm{\,}}\ensuremath{\mathrm{#1}}}
\newcommand{\eV}{\mathrm{eV}}
\newcommand{\MeV}{\mathrm{M}\eV}

\newcommand{\Mum}{\ensuremath{\umu}\ensuremath{\mathrm{m}}}
\newcommand{\mum}{\textrm{\,\ensuremath{\mathrm{\Mum}}}}
\renewcommand{\eqref}[1]{(\ref{#1})}
\newcommand{\com}{}

\mathchardef\ordinarycolon\mathcode`\:
\mathcode`\:=\string"8000
\begingroup \catcode`\:=\active
  \gdef:{\mathrel{\mathop\ordinarycolon}}
\endgroup

\begin{document}

\title{Optimized laser ion acceleration at the relativistic critical density surface} 

\author{Ilja Göthel}
\affiliation{Helmholtz-Zentrum Dresden-Rossendorf, Bautzner Landstra\ss e 400, 01328 Dresden, Germany}
\affiliation{Technische Universität Dresden, 01069 Dresden, Germany}
\author{Constantin Bernert}
\affiliation{Helmholtz-Zentrum Dresden-Rossendorf, Bautzner Landstra\ss e 400, 01328 Dresden, Germany}
\affiliation{Technische Universität Dresden, 01069 Dresden, Germany}
\author{Michael Bussmann}
\affiliation{Center for Advanced Systems Understanding (CASUS), Untermarkt 20, 02826 Görlitz, Germany}
\author{Marco Garten}
\affiliation{Helmholtz-Zentrum Dresden-Rossendorf, Bautzner Landstra\ss e 400, 01328 Dresden, Germany}
\author{Thomas Miethlinger}
\affiliation{Helmholtz-Zentrum Dresden-Rossendorf, Bautzner Landstra\ss e 400, 01328 Dresden, Germany}
\affiliation{Technische Universität Dresden, 01069 Dresden, Germany}
\author{Martin Rehwald}
\affiliation{Helmholtz-Zentrum Dresden-Rossendorf, Bautzner Landstra\ss e 400, 01328 Dresden, Germany}
\affiliation{Technische Universität Dresden, 01069 Dresden, Germany}
\author{Karl Zeil}
\affiliation{Helmholtz-Zentrum Dresden-Rossendorf, Bautzner Landstra\ss e 400, 01328 Dresden, Germany}
\author{Tim Ziegler}
\affiliation{Helmholtz-Zentrum Dresden-Rossendorf, Bautzner Landstra\ss e 400, 01328 Dresden, Germany}
\affiliation{Technische Universität Dresden, 01069 Dresden, Germany}
\author{Thomas E. Cowan}
\affiliation{Helmholtz-Zentrum Dresden-Rossendorf, Bautzner Landstra\ss e 400, 01328 Dresden, Germany}
\affiliation{Technische Universität Dresden, 01069 Dresden, Germany}
\author{Ulrich Schramm}
\affiliation{Helmholtz-Zentrum Dresden-Rossendorf, Bautzner Landstra\ss e 400, 01328 Dresden, Germany}
\affiliation{Technische Universität Dresden, 01069 Dresden, Germany}
\author{Thomas Kluge}
\email{t.kluge@hzdr.de}
\affiliation{Helmholtz-Zentrum Dresden-Rossendorf, Bautzner Landstra\ss e 400, 01328 Dresden, Germany}

\date{\today}

\begin{abstract}
In the effort of achieving high-energetic ion beams from the interaction of ultrashort laser pulses with a plasma, volumetric acceleration mechanisms beyond Target Normal Sheath Acceleration have gained attention.
A relativisticly intense laser can turn a near critical density plasma slowly transparent, facilitating a synchronized acceleration of ions at the moving relativistic critical density front. While simulations promise extremely high ion energies in in this regime, the challenge resides in the realization of a synchronized movement of the ultra-relativistic laser pulse ($a_0\gtrsim 30$) driven reflective relativistic electron front and the fastest ions, which imposes a narrow parameter range on the laser and plasma parameters.  We present an analytic model for the relevant processes, confirmed by a broad parameter simulation study in 1D- and 3D-geometry. By tayloring the pulse length and plasma density profile at the front side, we can optimize the proton acceleration performance and extend the regions in parameter space of efficient ion acceleration at the relativistic relativistic density surface.
\end{abstract}
\maketitle 

The acceleration of ions using lasers has great potential for breakthrough applications which require short pulse durations, high currents, high emittances or a small accelerator footprint~\cite{Bulanov2002,Roth2001,Fortney2009}. 
One of the most widely used performance benchmarks in practice is the maximum achievable ion energy.
In the pursuit for higher maximum ion energies, progress has been made with more powerful petawatt lasers~\cite{Schramm2017, Danson2019}, controlled pulse delivery~\cite{Obst-Huebl2018,Brack2020,Ziegler2020,Huebl2020} and optimized targets~\cite{Gaillard2011,Hilz2018,Tatomirescu2019}. 
Over the last two decades, the field has witnessed an improvement of the maximum proton energy from $58\unit{\MeV}$, obtained early in 1999\cite{Snavely-IntenseProtonBeams}, to recently more than $90\unit{\MeV}$\cite{Higginson2018,Hornung2020}. 
Despite the theoretical and experimental understanding that enabled the optimization of laser ion acceleration in recent years, it is still a long path towards the $\approx 200-300\unit{AMeV}$ required to enable applications such as tumor therapy~\cite{Bulanov2002,Kraft2010,Karsch2017}. \\
The above cited experiments\cite{Higginson2018,Hornung2020} were performed on relatively long pulse lasers with up to half a picosecond pulse duration. 
Those lasers are limited to low repetition rates, while ultra-short pulse lasers in the few tens of femtoseconds range can operate in the $10\unit{Hz}$ regime and are also more favourable due to reduced costs. 
Most notably, those machines can now be operated in a very reproducible and reliable fashion, such that, over the timespan of several months, protons with $~60\unit{MeV}$ can be generated routinely~\cite{Ziegler2020}. 

It is notable that virtually all experiments that report high proton energies were employing one variant of the established target normal sheath acceleration (TNSA)~\cite{Maksimchuck2000,Hatchett-ElectronPhotonAndIons,wilks} or transparency enhanced TNSA method~\cite{DHumieres2005,Yan2010,Mishra2018,Poole2018}. 
Hence, it is clear that in order to soon reach the envisioned $200\unit{AMeV}$ or more, breakthrough innovations are needed. 
The electrons act as a mediating medium in the energy transfer process between the laser and the ions, and thus mitigate the potential efficiency of the whole acceleration process. 
One major drawback of TNSA is that the ions from the solid target rear surface are accelerated indirectly via the electrons by expansion of the plasma anode. 
As a consequence, the light protons only briefly experience the maximum possible acceleration field at the rear surface before they detach from the bulk, the target surface degrades, and the electron sheath explodes transversely. 
Candidates to overcome this limitation are collisionless shock acceleration (CSA) and radiation pressure acceleration (RPA). 
There, the laser pushes an electron sheath in front of the ions. 
This can happen either at the front surface of a foil (hole boring RPA\com{, HB-RPA}~\cite{Esirkepov-LaserPiston,Robinson2009}), or it pushes forward all electrons of a very thin foil. 
In the latter case, the laser pressure continuously accelerates the whole foil and hence the ions (light sail RPA\com{, LS-RPA}~\cite{Simmons1993,macchi2009light}). 
However, transverse instabilities and subsequent plasma heating pose practical limitations on the acceleration of ions to high energies that are challenging to overcome~\cite{Pegoraro2007,Palmer2012}. 

Another particularly interesting alternative are near critical density targets, e.g. hydrogen jet targets. 
Those target systems can provide for cylindrical and planar micron sized geometry and virtually debris-free high repetition rate operation\cite{Obst2017,Curry2020}. 
One theoretical mechanism of continuous and direct energy transfer from the laser to ions is a form of relativistic hole boring\com{: 
Instead of boring into the plasma, the laser now penetrates into the relativistically transparent plasma as it pushes forward the relativistic critical density front, synchronized to the acceleration of ions. 
This is the mechanism of ''synchronized acceleration of ions using slow light'' described in~\cite{Brantov2016a,Bychenkov2016,Brantov2017a,Brantov2017b,Brantov2017c}. 
However, the naming can be confusing as the core of the mechanism is not the change of the propagation velocity of the laser light inside the plasma but rather the fact that the point of laser reflection, i.e. where the ions are accelerated, penetrates \emph{slowly} into the plasma. } \\
\com{The laser penetration happens progressively as }the laser turns the plasma partially transparent at the front, by virtue of relativistic mass increase of the electrons oscillating (linear polarization) or circulating (circular polarization) in the intense laser field. 
The front that separates the relativisticly transparent region from the opaque plasma with density above the relativistic critical density (\com{below} referred to as ”relativistic \com{transparency} front”\com{, RTF}) moves forward slowly (compared to the speed of light) and accelerates as the laser intensity increases, which gives the ions time to accelerate and follow it in a synchronized fashion.
The acceleration force is generated \com{via} the charge separation field induced by the laser pressure at the front. \\
\begin{figure}
    \centering
    \includegraphics[width=\linewidth]{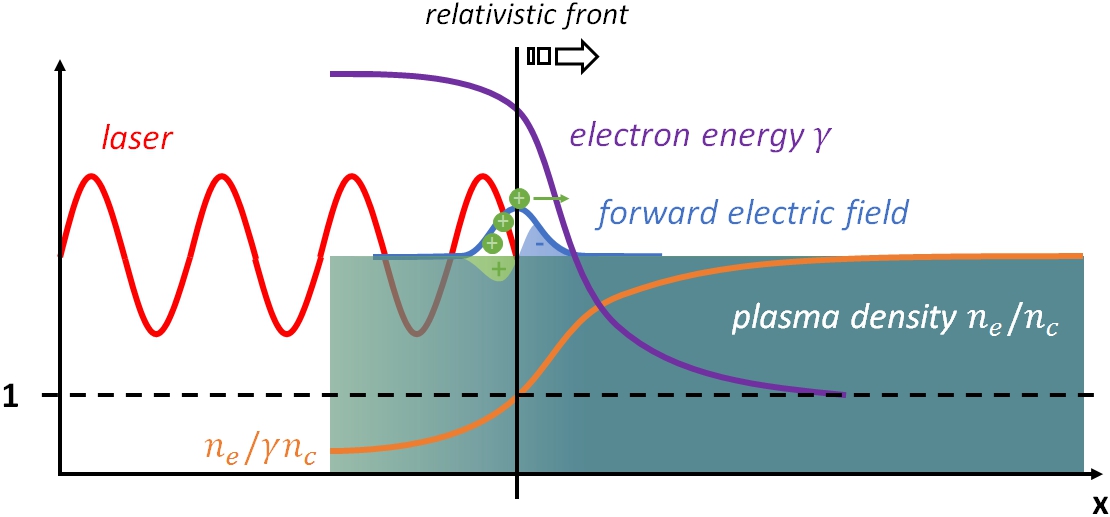}
    \caption{Schematics of the \com{RTF-RPA} mechanism.}
    \label{fig:scheme}
\end{figure}
\com{In this respect, this mechanism is similar to hole boring RPA, only here instead of the critical density surface at the target front the radiation pressure is now exerted at the relativistic transparency front inside the plasma. 
Therefore, we call this process \emph{\com{RTF-RPA}} throughout this paper. }
Generally, the velocity at which the laser bores into the plasma depends on the plasma density and laser intensity; the equilibrium \com{RTF} velocity increases with intensity and decreasing electron density~\cite{Liu2020}. 
Simulations could demonstrate that in the optimal case \com{when the laser acceleration of the \com{RTF} is synchronized to the ion acceleration,} the ion energy can be dramatically increased \com{compared} to, e.g., hole-boring or TNSA~\cite{Brantov2016a,Bychenkov2016,Brantov2017a,Brantov2017b,Brantov2017c}. 

Here we develop a self-consistent semi-analytic model to calculate the maximum ion energies in the \com{RTF-RPA} regime and use this to infer optima for the laser pulse duration and plasma density profile. 

\section{RTF-RPA mechanism}
Before we introduce the semi-analytic model, we briefly review the \com{RTF-RPA} mechanism. 
Fig.~\ref{fig:scheme} shows a schematic picture of \com{RTF-RPA}. 
The laser penetrates the plasma with a free electron density $n_e$, \com{accelerating or} heating \com{them} at the front and thereby increasing the critical density $\gamma n_c$. 
At the \com{RTF} ($n/\gamma n_c = 1$) the laser is reflected and a charge separation field accelerates the ions. 
As mentioned above, \com{RTF-RPA} shares certain similarities with hole boring: The reflective surface where $\gamma n_c = n_0$ is pushed forward and ions are reflected at the moving surface. 
The characteristic difference is that with \com{RTF-RPA} it is not the plasma material $n_0$ itself that is pushed forward but the reflective surface moves via a progressive increase of $\gamma$. 
In this case the penetration velocity is accelerating much faster with an increase of the laser intensity so that the front can keep up with the reflected ions. 
Perfect synchronization is not required, as long as the \com{RTF} is accelerated fast enough for the ions to be caught up and accelerated repeatedly. 

\begin{figure}
    \centering
    \includegraphics[width=\linewidth]{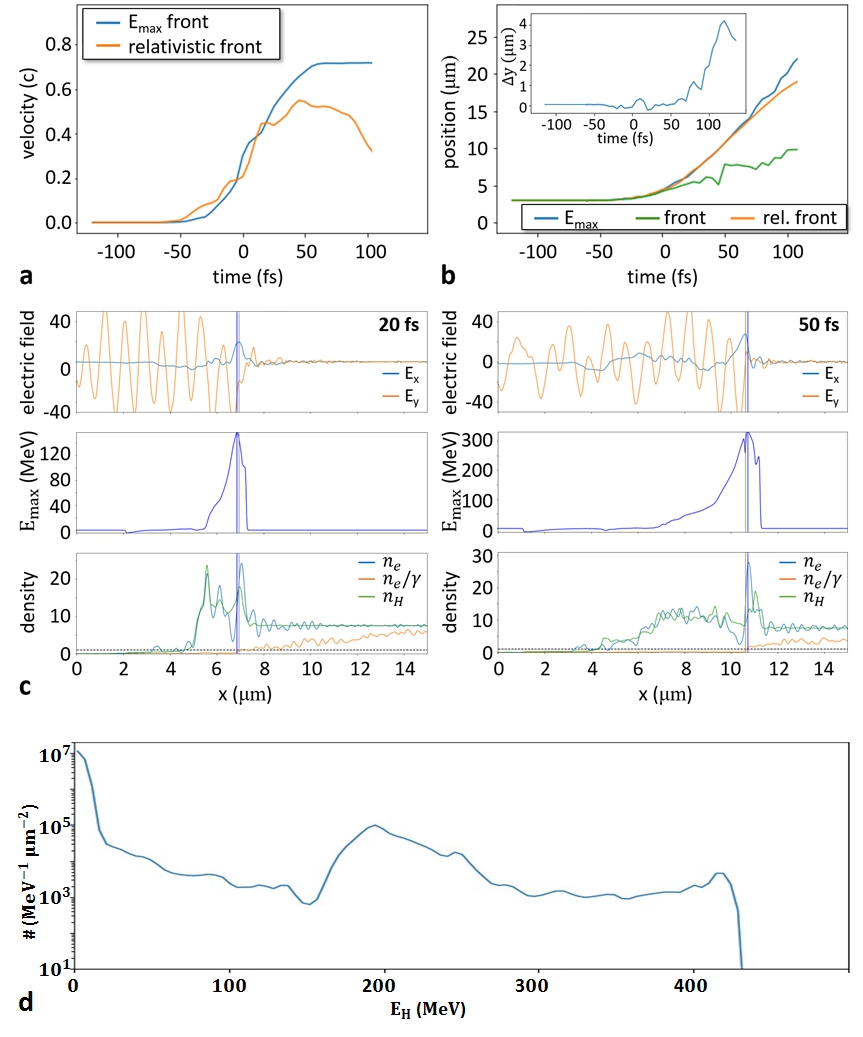}
    \caption{Velocity (a) and position (b) of the relativistic density front and highest energy ions as function of time. (c) shows the electric field (top), maximum ion energy (center) and densities (bottom) for two timesteps during the \com{RTF-RPA} process, the position of the relativistic density front is marked by the blue vertical line. (d) Proton spectrum at the end of the simulation. Simulation parameters: see main text.}
    \label{fig:1}
\end{figure}
In Fig.~\ref{fig:1} we exemplify this for an intense laser with $a_0=50$, linear polarization $P=1$, and $t_\mathrm{FWHM}=40\unit{fs}$ FWHM pulse duration for a Gaussian temporal intensity profile $a(x,t)^2$. 
The plasma density was optimized w.r.t. maximum ion energy gain by \com{RTF-RPA} to $n_0 \approx 8 n_\mathrm{c}$ in a 1D geometry SMILEI~\cite{Derouillat2018} simulation. 
Since \com{RTF-RPA} is essentially a 1D process, we expect the 1D simulations and a 1D model to capture the essential physics. 
Later we will compare our results also to 3D simulations, which in deed show the same characteristic features and trends.

The simulated plasma consists of free protons and electrons. 
It starts at $3.75\lambda_\mathrm{L}$ from the left simulation box boundary and extends through the whole box to the right boundary. 
The simulation duration and box size were chosen so that the laser while it is being reflected and the fastest electrons traveling at speed of light do not reach the right box boundary, thereby ignoring any effects a rear surface might induce. 
The resolution was set to $\Delta x=\Delta y =\Delta t \cdot \sqrt{2}/0.99 =\lambda_\mathrm{L}/16$. 
Here and in the following, we use the conventional dimensionless quantities by setting $c=e=m_\mathrm{e}=\omega_\mathrm{L}=n_c=1$. 
For simplicity we show our results in SI units, where we assumed $\lambda_\mathrm{L}=0.8\unit{\mum}$. 
The full simulation parameters as well as all relevant scripts necessary to reproduce the results is available online~\cite{goethel2021scripts}. 

One can see the \com{RTF-RPA} mechanism accelerating ions to very high energies. 
Panels a and b show the position and velocity of the fastest ion (blue) and of the laser reflection (\com{RTF}, orange), respectively. 
As a reference, in panel b we also plot the position of the (non relativistic) critical density surface (green). 
While the latter can be seen to move only slowly as a result of the radiation pressure, the point of reflection quickly penetrates due to relativistic transparency into the plasma, in sync with the fastest ions. 
Even though initially the ions are accelerating slower than the \com{RTF}, later the \com{RTF} accelerates slowly enough so that the ions start to catch up and co-move with the accelerating field essentially during the whole laser upward ramp. 
The density and field distribution are shown in c. 
In the lower panel, the electron, ion and relativistically corrected electron density profiles $n_\mathrm{e}$, $n_\mathrm{H}$, $n_\mathrm{e}/\gamma$ are plotted normalized to $n_\mathrm{c}$ for two times, 20 fs and 50 fs after the laser maximum arrival at the target front. 
The laser is seen to penetrate into the plasma, where it is reflected at the point where the relativistically corrected density surpasses the critical density $n_\mathrm{c}$.  
This point is marked by a blue vertical bar also in the middle and top insets, that show the maximum ion energy profile and electric field profiles respectively. 
The longitudinal electric field profile shows a distinct peak around the laser reflection point due to the charge separation, and the position of the maximum ion energy coincides with that position. 
Specifically, for a plasma density of $n_0 \approx 8 n_\mathrm{c}$, the \com{RTF-RPA} process works best for the specific laser pulse duration and peak intensity of the simulation shown in Fig.~\ref{fig:1}, where the most energetic ions can co-move with the \com{RTF} almost during the whole laser pulse duration. 
The protons eventually reach a maximum kinetic energy exceeding $400\unit{MeV}$, compared to only $75\unit{MeV}$ expected for relativistic hole boring~\cite{Robinson2009}. 

In fact, hole boring \emph{is} observed for higher plasma density, as will later be shown in the context of Fig.~\ref{fig:optT} and \ref{fig:6}c. 
For lower densities, the plasma is transparent and the laser induced charge separation field only sweeps briefly over the ions. 
Consequently, the ion energy drops dramatically, rendering the energy gain negligible in this regime. 
For even smaller plasma density or at high values of $a_0$, another regime of efficient ion acceleration in a plasma wakefield is predicted~\cite{Liu2016,Liu2018}, which is beyond the scope of this paper, though it becomes visible at the lowest densities and highest laser intensity shown later in Fig.~\ref{fig:optT}. 
We note that the absolute ion energy in the \com{RTF-RPA} observed in our simulations is in good agreement with the results in~\cite{Brantov2016a}, where for the same laser parameters a 3D simulation has shown proton acceleration in a $\mathrm{CH_2}$ foil to $350\unit{\MeV}$ by \com{RTF-RPA}, albeit at somewhat higher density of $n_0=20 n_\mathrm{c}$~\footnote{This shift of the peak is (at least partially) explained by two major differences between the simulated system in 1D and 3D, that affect the \com{RTF-RPA} process, though it is inherently a 1D phenomenon. With added transversal dimensions, the relativistic density can be reduced not just by electron relativistic mass increase, but also number density decrease due to the transverse ponderomotive force; also, self focusing occurs that can increase $a_0$ quite significantly, which in turn also shifts the peak to higher bulk densities.}. \\
 
The underlying ultimate limitation for the maximum ion energy at a given pulse duration and peak intensity is the coupling of two variables with the one parameter of plasma density: \com{RTF} acceleration and maximum \com{RTF} velocity. 
The former one needs to be matched to that of the ions so that they can stay close to the \com{RTF}, and the latter should be as high as possible as it dictates the maximum possible ion energy limit. 
The optimum plasma density consequently resembles a compromise between those two optimization goals, since an increase of the \com{RTF} velocity would inevitably leave the ions behind during the initial acceleration phase. 

\section{Results}
In order to increase the ion energies achievable with \com{RTF-RPA}, ideally one would like to reduce the plasma density in order to increase the terminal \com{RTF} velocity, while at the same time keeping the acceleration slow and synchronized to the ions. 
There are two principal routes to achieve this goal: modifying the pulse shape or modifying the plasma density. 
We first will examine how a simple laser pulse duration and density variation can optimize the ion energies for a given laser intensity, and then optimize the ion acceleration by introducing a compression layer in front of the plasma. Its dramatic effects of increase of the maximum ion energy and larger range of lower laser intensities where \com{RTF-RPA} can still be active are the main findings of this paper. 
\subsection{RTF-RPA analytic model}
Before looking at the details of the simulation results, we briefly review the one dimensional \com{RTF-RPA} process. 
For small values of $a_0$ the laser pushes forward the front surface via HB-RPA. 
For larger $a_0$ the laser can penetrate into the plasma by turning the plasma relativistically transparent at the front. 
The velocity of the point of reflection of the laser can be modeled by
\begin{equation}
    \beta_\mathrm{f}=\begin{cases}
               \beta_\mathrm{HB} & \text{if } v_\mathrm{HB} > v_\mathrm{S} \\
               \beta_\mathrm{S} & \text{if }  v_\mathrm{S} > v_\mathrm{HB}
               \end{cases},
    \label{eqn:Liu}
\end{equation}
where $\beta_\mathrm{HB}=\frac{\Sigma}{1+\Sigma}$ is the HB velocity with $\Sigma=\sqrt{\frac{Pa^2}{2 m_\mathrm{i} n_\mathrm{i}}}$~\cite{Robinson2009}, and \begin{equation}
 \beta_\mathrm{S} = K_+ - K_- -1
 \label{eqn:v_front}
\end{equation}
is the velocity of the relativistic critical density front with
\begin{equation}
    K_\pm(a,n_\mathrm{e})=\left[\frac{a(x_\mathrm{f},t)}{b(n_\mathrm{e})} \left(\sqrt{1+\frac{a(x_\mathrm{f},t)}{27 b(n_\mathrm{e})}}\pm 1\right)\right]^{1/3}
 \label{eqn:v_front_K}
\end{equation}
and $b=\pi^2 P n_\mathrm{e}/16$~\cite{Liu2020}. 
The lower subscript $f$ refers to quantities of the \com{RTF} where the laser is reflected. The coefficient $P$ refers to the polarization, see below. 
The plasma is assumed to be homogeneous with electron density $n_\mathrm{e} = Z n_\mathrm{i}$. 
Here we introduced the empiric factor $P$ to approximate the acceleration of the front velocity for both linear laser polarization ($P=1$) and circular polarized light ($P=2$). 
For circular polarized light Eqn.~\ref{eqn:v_front} and~\ref{eqn:v_front_K} were found to be in good agreement with numerical simulations~\cite{Liu2020}. 
Here, we find that also for linear polarization we obtain a good agreement of the model with our simulations when introducing the factor $P$, cp. Fig.~\ref{fig:1}~\footnote{Naively, this an intuitive explanation: The plasma in the downstream region is heated due to accelerated electrons. Far from the \com{RTF} this effect vanishes and $n_0/(T_{e,\infty}+1)\approx n_0$, while directly at the front $n_0/(T_{e,0}+1)\approx 0$. 
The laser has to push against this downstream gradient, i.e. has to work against the average $n_0/2$, hence  $P\equiv 2/(T_{e}+1)\approx 1$. The mathematical derivation can be found in the appendix. }. 
For $a\gg b$ the above equation can be approximated to the simple form\cite{Liu2020}
\begin{equation}
    \beta_\mathrm{f}\approx \frac{1}{\sqrt{1+P\pi^2n_0/a}}.
    \label{eqn:beta_limit}
\end{equation}\\
For the case shown in Fig.~\ref{fig:1} with $a_0=50$ and $n_\mathrm{e}=8$ the predicted velocity of the \com{RTF} from Eqn.~\eqref{eqn:v_front} is $\beta_\mathrm{f}=0.6$, which compares well to a maximum value of $\beta_\mathrm{f}^\mathrm{PIC}=0.57$ in the simulation. \\
Upon laser reflection a charge separation field is set up at the \com{RTF} which can reflect downstream ions with peak field strength~\cite{Liu2020} \begin{equation}
    \hat E_x(t)=\sqrt{2P\frac{1-\beta_\mathrm{f}(t)}{1+\beta_\mathrm{f}(t)}}a(x_\mathrm{f}(t),t).
    \label{eqn:Ep}
\end{equation}
While the laser pulse intensity ramps up, the velocity at which the laser turns the plasma transparent increases, i.e. the \com{RTF} accelerates and can catch up with the ions. 
The charge separation field at the front then reflects and further accelerates them repeatedly. 

We now combine the above in order to obtain an analytic closed description of the \com{RTF-RPA} process in 1D. 
The acceleration of the ions $dp/dt$ during the laser pulse irradiation can be calculated from the usual HB velocity as long as $\beta_\mathrm{HB}<\beta_\mathrm{S}$, and later for larger laser strengths $a(x,t)$ by the Equation of motion (EOM) with 
\begin{equation}
    \frac{dp}{dt}= Z E_x(x_i,t). 
    \label{eqn:EoM}
\end{equation}
We approximate $E_x$ to be spatially constant around the front position, $E_x(t)\cong \hat E_{x}(x_\mathrm{f}(t),t)$ if $\left|x-x_\mathrm{f}\right| < L/2$, and $E_x=0$ otherwise. 
We find empirically from our simulations shown below that the characteristic length $L$ over which the accelerating field extends can be well approximated by a quarter plasma wavelength in the upstream region, \begin{equation}
    L\approx \lambda_\mathrm{p}/4\approx \lambda_\mathrm{L}/\left(4\sqrt{n_0/\gamma}\right)
    \label{eqn:L}
\end{equation}
where $\gamma\approx \sqrt{1+a_0^2/2}$ is the average Lorentz factor of the electrons in the upstream region. 

The EOM can be easily integrated numerically to obtain the ion velocity and energy. 
For example, we can derive the optimal values for the pulse duration, density or laser amplitude with respect to the final ion energy, see the following sections. \\
In the optimum case with the maximum ion energy for a given laser, the EOM can easily be integrated analytically. 
The maximum possible ion energy is dictated by the maximum velocity of the \com{RTF} $\beta_\mathrm{f,max}$, which is a function of $a_0$ and $n_\mathrm{e}$. 
The local optimum pulse durations and plasma densities are given when the accelerated ions are reflected the last time exactly when the laser maximum reaches the \com{RTF}. 
Then, they whiteness the full electrostatic potential around the \com{RTF} at its maximum strength.  \\
For these optimum cases we start integrating the EOM after the ions have obtained the maximum \com{RTF} velocity with respective energy $m_i\gamma_0$, $\gamma_0=\left(1-\beta_\mathrm{f,max}^2\right)^{-1/2}$. 
\begin{figure}
    \centering
    \includegraphics[width=\linewidth]{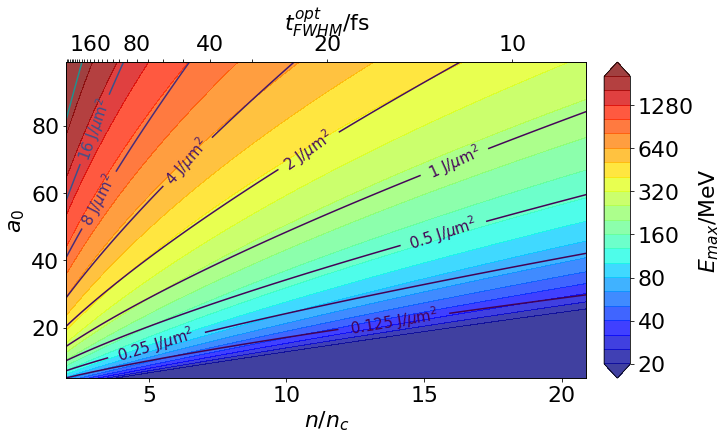}
    \caption{Maximum proton energies as a function of plasma density and laser strength obtained from Eq.~\eqref{eqn:max} for $P=1$ and $\lambda=0.8\unit{\mum}$. The top axis shows the corresponding $t_{FWHM}^{opt}$ from Eqn.~\eqref{eqn:empiricOptimum} (i.e. the detailed numerical integration of the EOM~\eqref{eqn:EoM} in Fig.~\ref{fig:optT}). The respective laser pulse energy densities $a_0^2 t_{FWHM}^{opt}$ are shown by iso-contour lines for reference. }
    \label{fig:maxEnergy}
\end{figure}
We make a variable transformation
\begin{equation}
    \frac{1}{dt}=\left(1-\beta_\mathrm{f}\right)\frac{1}{d\phi}
\end{equation}
and then integrate from $\phi_0=0$ to $\phi_\mathrm{max}=L$. 
We assume $E_{x}(t)=E_0$ is constant to first order around the laser maximum during the short post-acceleration phase. 
After some algebra we obtain
\begin{equation}
    \gamma_\mathrm{max}=\sqrt{\frac{A^2}{\left(1-\beta_\mathrm{f,max}\right)^2}+1}
    \label{eqn:max}
\end{equation}
with $A=\gamma_0 \beta_\mathrm{f,max} \left( 1 - \beta_\mathrm{f,max} \right) + Z E_0 L / M_i$. 
The optimum maximum energies of protons are plotted in Fig.~\ref{fig:maxEnergy} as a function of laser strength $a_0$ and plasma density $n_0$. 

Generally, the final ion energy is sensitive to the exact position in the longitudinal field behind the \com{RTF} at the time of the laser peak arrival. 
In these cases the EOM needs to be integrated numerically piecewise for each reflection in order to obtain the position of the fastest ions ions in the charge separation sheath. 

\begin{figure*}
    \centering
    \includegraphics[width=\linewidth]{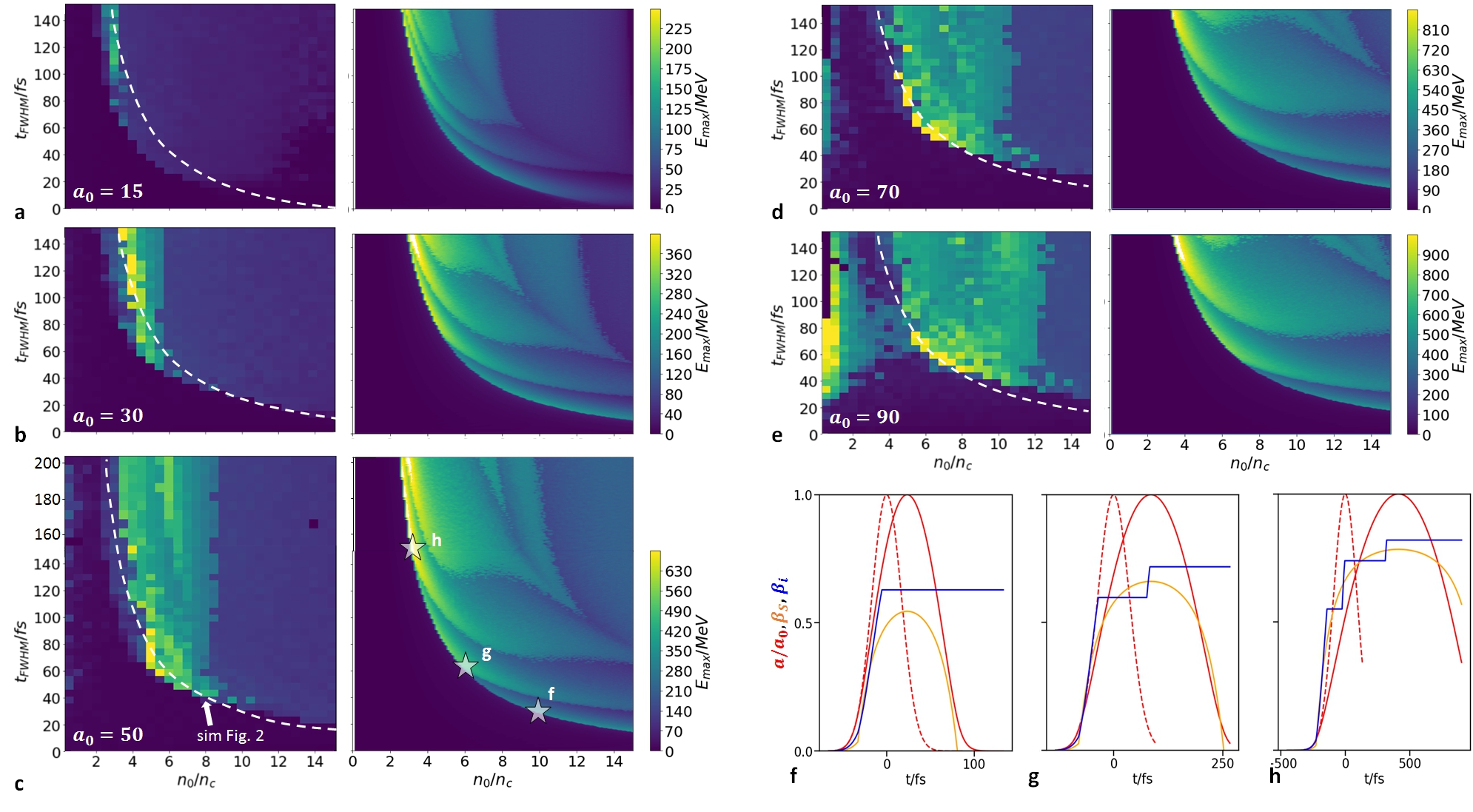}
    \caption{Maximum proton energies as a function of pulse duration $t_\mathrm{FWHM}$ and plasma density $n_0$ for a Gaussian laser pulse temporal profile and homogeneous density $n_0$. Numerical integration of the semi-analytical model on the right, SMILEI 1d simulations on the left. White dashed lines mark the optimum calculated with the semi-analytical model.  Panels f-h correspond to the stars in (c, right) and show the velocity of the front (orange) and fastest ions (blue), and the normalized laser field strength $a(x=0,t)/a_0$ (red, dashed) and $a(x_\mathrm{f}(t),t)/a_0$ (red, solid).}
    \label{fig:optT}
\end{figure*}
\subsection{Optimum pulse duration}
The pulse duration at which \com{RTF-RPA} works is a function of laser amplitude $a_0$ and plasma density $n_0$. 
For laser pulses too short, the ions cannot keep up during the ramp-up, for pulses too long the ions energy will depend on the exact timing of the \com{RTF} dynamics and ion acceleration dynamics. 
To find the optimum pulse duration for specific values of $a_0$ and plasma density $n_0$, we selected 5 exemplary values for $a_0$ and varied the plasma density and laser pulse duration. 
For each set we integrate the EOM and obtain the final hydrogen kinetic energy, shown in the right columns of Fig.~\ref{fig:optT}a-e. 
Above, in Fig.~\ref{fig:1}, we have shown a simulation for $a_0=50$ and $n_0=8 n_\mathrm{c}$. 
Intuitively, at constant $a_0$ one would expect the ion energy to increase for smaller plasma densities, since the laser front can then propagate faster and the co-moving ions reach a higher final energy. 
What we find from the numerical solutions of the EOM is in fact such an increase, but only for the envelope along the line of the respective optimum pulse durations. 
Only at certain values of the plasma density the highest ion energies expected from Eqn.~\eqref{eqn:max} are obtained. 
Those local maxima resemble those optimum conditions when the ions are reflected exactly at the time when the laser maximum reaches the \com{RTF} and for which Eqn.~\eqref{eqn:max} was derived. 
This can be seen in panels f-h. 
In panel f the ion are only reflected once, but the density is smaller than optimum so that the ions are reflected before the laser maximum has reached. 
Panel g and h show the respective optimum cases for two and three reflections. 
Hence, each \textit{leaf}-shaped area in the $n_\mathrm{e}-t_\mathrm{FWHM}$-space corresponds to a specific amount of reflections. 
When the reflection occurs exactly at the laser maximum arrival, the local optima in the respective leaf w.r.t. ion maximum energy is observed. 

To confirm those results we performed a series of 1D PIC simulations with varying initial plasma densities and pulse durations at different peak laser strengths.
The results are plotted in Fig.~\ref{fig:optT} for the different laser strengths. 
We extracted the proton energies at $140\unit{fs}$ after laser peak arrival on target, when the laser pulse has passed.
For a fixed pulse duration, in our parameter range the maximum proton energies due to \com{RTF-RPA} are generally obtained at the respective smallest plasma density at which \com{RTF-RPA} is still active (onset of \com{RTF-RPA}, white dashed lines). Only at the highest simulated laser strength we observe similarly high ion energies by ion wave breaking~\cite{Liu2016,Liu2018}) at smaller density, which is beyond the scope of this paper. 
The optimum combination of pulse duration and plasma density is not changing much with laser strength.
It roughly follows the empiric conjecture
\begin{equation}
    t^\mathrm{opt}_\mathrm{FWHM} \approx m_i \left(P n_0\right)^{-3/2},
    \label{eqn:empiricOptimum}
\end{equation}
and is in very good agreement with the semi-analytic model, despite its simplistic assumptions. 
The exact position of the local optimum along the line given by this equation is quite sensitive to the exact timing of when the ions are reflected the last time, which translates into narrow regions of high proton energy. 

\begin{figure}
    \centering
    \includegraphics[width=\linewidth]{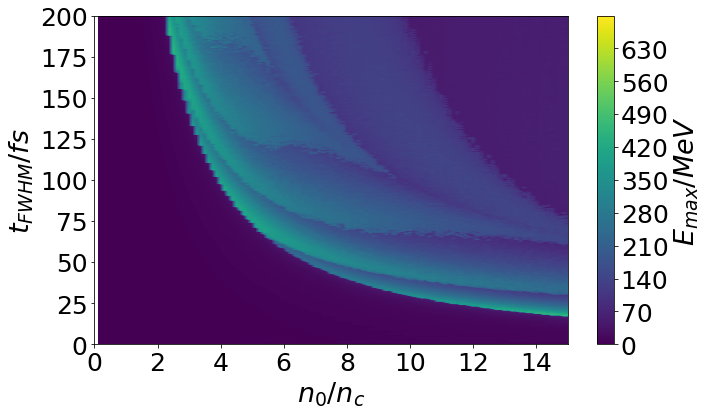}
    \caption{Semi-analytical model: Proton energy as a function of plasma density and laser pulse duration for constant pulse energy, equal to that used in Fig.~\ref{fig:1}. The leaves clearly each maximize the energy at the same proton energy level, hence the maximum proton energy has no significant dependence on pulse duration. }
    \label{fig:Econst}
\end{figure}

One can combine Eqn.~\eqref{eqn:empiricOptimum} and \eqref{eqn:max} to obtain the maximum ion energy as function of $a_0$ and $t_{FWHM}^{opt}$, cp. Fig.~\ref{fig:maxEnergy} top horizontal axis. 
In the range between $a_0>n_0$ and $a_0\lesssim 100$ one obtains the approximate scaling of $E_{max}\propto a_0^{3/2}/n_0$ at constant $t_{FWHM}$. 
We can also evaluate the maximum ion energy as a function of laser pulse energy density (i.e. $a_0^2 t$), as indicated by the solid lines in Fig.~\ref{fig:maxEnergy}. 
For a given fixed laser pulse energy density between $0.5$ and $4\unit{J/\mum}^2$ we find that the maximum possible ion energy is expected to be approximately independent of the laser intensity and pulse duration. 
In Fig.~\ref{fig:Econst} the results of the numeric integration of the EOM are shown for constant laser pulse energy density matching that of the reference simulation in Fig.~\ref{fig:1} ($\approx 1.7\unit{J/\mum}^2$) and one can readily see that in deed the the optimum maximum ion energies at the local maxima do not change much with the pulse duration. 
For much larger laser pulse energies longer pulse durations would be favoured while for much smaller energies shorter pulse durations are more favourable. 


However, for very short pulse durations the synchronization between ions and the laser needs to be increasingly more precise, because the difference between too fast laser propagation (ions are left behind early in the laser upramp) and too slow laser propagation (ions are pushed too far ahead and the laser cannot catch up during the short pulse duration) becomes progressively smaller. 
This means, that in practice the paramater range where \com{RTF-RPA} is accessible gets progressively smaller and in the simulations \com{RTF-RPA} does effectively not occur for short pulse durations or low laser intensities. 
Instead, at the predicted optimum \com{RTF-RPA} condition we observe a direct transition between transparency and hole boring, with no \com{RTF-RPA} in between. 
The threshold pulse duration below which no \com{RTF-RPA} can be observed in the simulations increases with decreasing laser strength. 
While for $a_0=90$ we can observe \com{RTF-RPA} down to almost $t_\mathrm{FWHM}\cong 30\unit{fs}$, for $a_0=15$ we can only see it down to $60\unit{fs}$. 

On the other hand, for longer pulse durations the density range at which we observe \com{RTF-RPA} increases due to the fact that the ions are pushed forward and reflected at the accelerating \com{RTF} multiple times. 
Slower laser intensity ramp-up in principle simply results in more reflections until the final energy is reached.
Additionally, the maximum ion energy becomes less sensitive to the exact timing of the last reflection due to the temporally spread out maximum of the laser intensity and front velocity.
\begin{figure*}
    \centering
    \includegraphics[width=\linewidth]{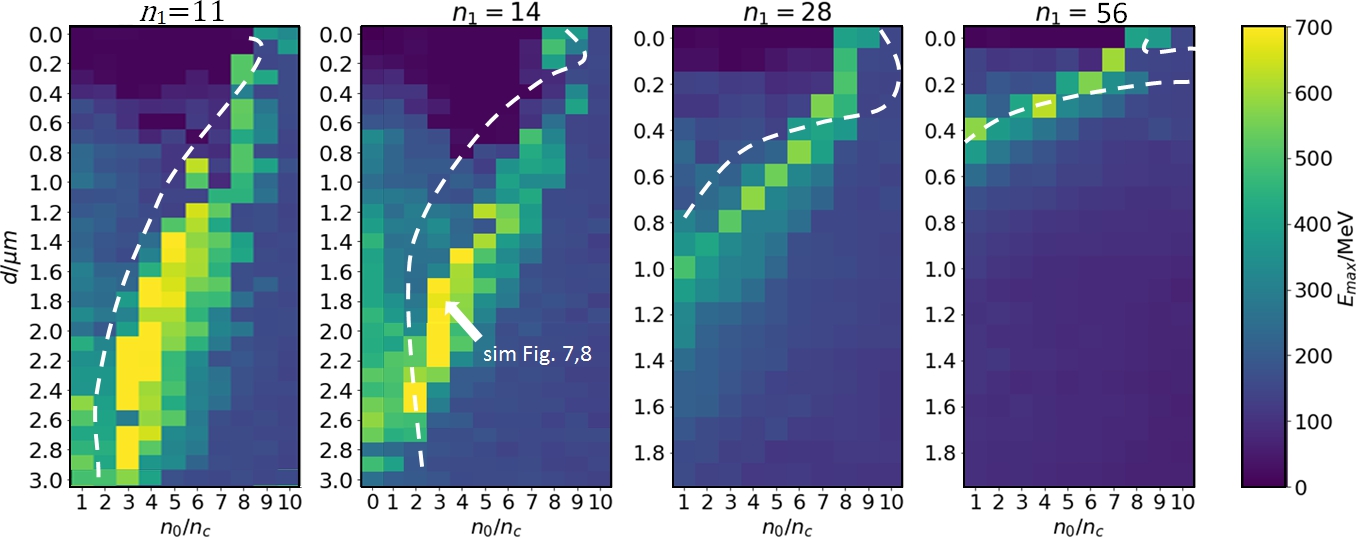}
    \caption{Maximum proton energy after the acceleration process for laser strength $a_0=50$ as a function of rear layer density $n_0$ and front layer thickness $d$ for front layer density $n_1/n_\mathrm{c}=11, 14, 28, 42$ (from left to right). The predictions of the semi-analytic model for the onset of \com{RTF-RPA} are indicated by the white dashed lines.}
    \label{fig:2}
\end{figure*}
Hence, it follows that the higher ion energies occurring at smaller plasma densities due to faster laser penetration require an increasingly longer pulse duration and hence higher laser pulse energy in order for the ions to keep up with the \com{RTF}.
However, at the longest pulse durations we observe considerable density steepening by the laser pulse at the front, which slows down the acceleration of the \com{RTF}. 
This leads to a dephasing of the laser with the ions and inhibits the synchronization with the acceleration ions eventually. 
We performed an extensive simulation study testing negative density gradients to circumvent this problem, but due to a very unstable, unpredictive onset of the steepening the synchronization is very delicate between steepening and transparency -- practically we could not achieve it in our simulations for long pulse durations. 

In practice, this means that \com{RTF-RPA} is limited both for small as well as high laser pulse energies, which poses an ultimate limit to Eqn.~\eqref{eqn:max} and the maximum achievable ion energy. 
We can therefore expect the existence of an optimum pulse duration. 
In the simulations this occurs at the parameter combination at which the second or third ion reflection happens at the laser pulse maximum. 
In this sense, the case shown in Fig.~\ref{fig:1} hence already resembles a close-to-optimum scenario for a Gaussian beam profile at the given pulse energy density. 

\subsection{Optimum density compression layer}
In order to go circumvent the limits set by hole boring and plasma density steepening on the feasible laser pulse durations, we need another approach to reach smaller plasma densities than just increasing the laser pulse duration. 
We propose an inhomogeneous target plasma density to match the velocity of the \com{RTF} to that of the accelerating ions. 
A higher density at the target front is employed to  achieve this goal by slowing the laser penetration during the initial phase. 
Specifically, in the following we study a two-layer target consisting of a thin dense front compression layer and a low density bulk rear. 
The goal is to set the bulk density to a much lower density than the optimum \com{RTF-RPA} density of a homogeneous density to facilitate a high terminal velocity of the \com{RTF}. 
The injection problem, i.e. acceleration rate matching of the ions, is managed by the dense front layer.
Experimentally, such a front layer could for example be realized by dedicated laser prepulse compression or a tailored laser pulse rising edge contrast profile.
Since this compression layer is of higher density than the remaining bulk, the laser needs some time to bore through.
This gives the ions enough time for the initial acceleration and to later keep up with the \com{RTF} once the laser bores through the front layer.
Then, the \com{RTF} accelerates further to the final velocity in the less dense rear and -- for the same laser parameters -- the comoving ions should reach much higher energies than in the case of respective optimum density homogeneous plasma. \\

\begin{figure}
    \centering
    \includegraphics[width=\linewidth]{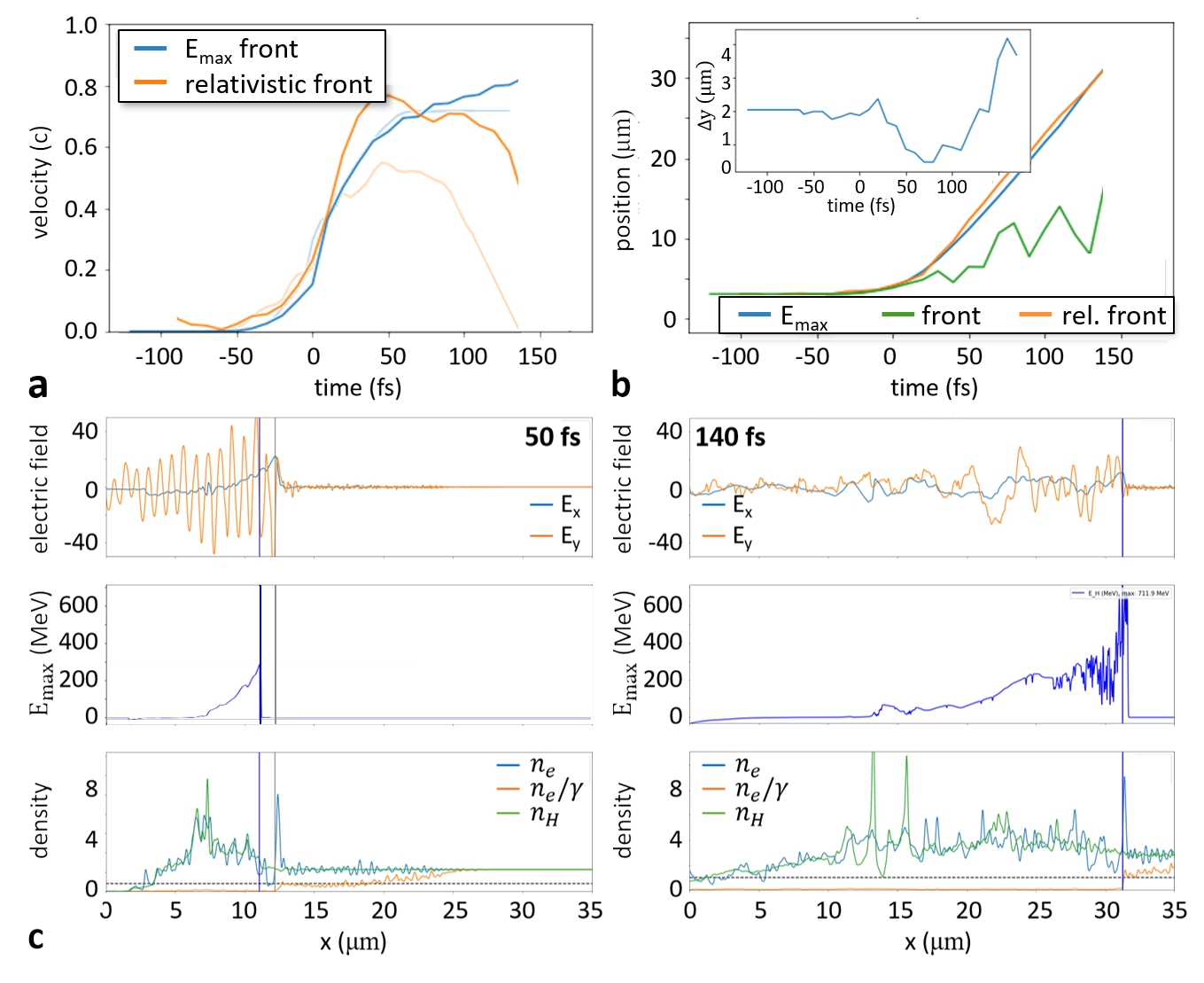}
    \caption{Plasma dynamics for the optimum front layer configuration at $a_0=50$, $n_0=3n_\mathrm{c}$, $n_1=14n_\mathrm{c}$, $d=1.8\unit{\mum}$. Light colors in panel a) show the optimum case without compression layer $d=0$ from Fig.~\ref{fig:1}.}
    \label{fig:3}
\end{figure}
The effectiveness of this method is summarized in a parameter scan in Fig.~\ref{fig:2}, again for the same laser parameters as in Fig.~\ref{fig:1} with $a_0=50$ and $40\unit{fs}$ FWHM pulse duration.
For a series of 1D PIC simulations of a plasma consisting purely of electrons and protons, we plot the maximum proton energy for different combinations of the front layer thickness and rear layer density. 
For panels a, b, c we defined the front compression layer density $n_1=14$, $28$, $56$, respectively. 
For finite values of the front layer thickness we find that the optimum density $n_0$ gets reduced. 
The optimum front layer thickness is shown by the white dashed lines, as calculated by numerically solving the EOM including the two stages of \com{RTF-RPA} first in the dense and secondly in the less dense plasma. 
It is again in good agreement with the simulations. 
The larger the front layer thickness $d$ and density $n_1$, the larger is the reduction of the optimum $n_0$ for a given $a_0$. 
At the same time, we can observe a substantial increase in ion maximum energy, as predicted. 
For fixed laser intensity and pulse duration the optimum two-layer plasma profile is found at $n_0\cong 3n_\mathrm{c}$, $n_1=14n_\mathrm{c}$, $d\cong 1.8\unit{\mum}$. 
Then the maximum ion energy jumps to approximately $800\unit{\MeV}$, which is twice the energy of the optimum at homogeneous plasma density using the same laser parameters. 
\begin{figure}
    \centering
    \includegraphics[width=6.2 cm]{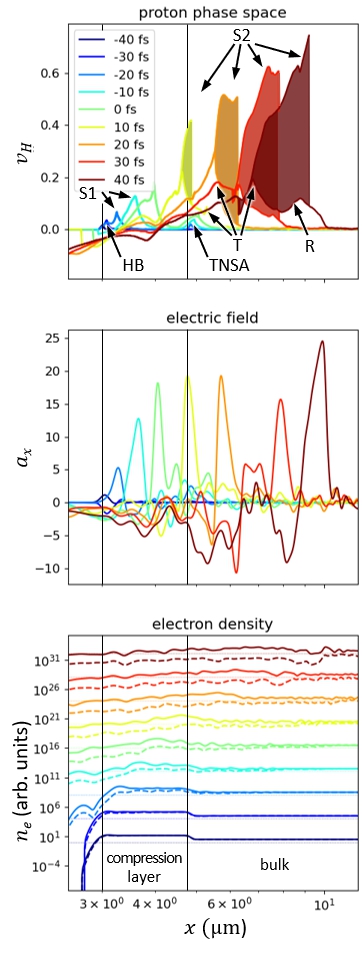}
    \caption{(a) Proton phase space, (b) accelerating electric field and (c) electron density $n_\mathrm{e}$ (solid) and relativistically corrected electron density $n_\mathrm{e}/\gamma$ (dashed) for various timesteps before and during \com{RTF-RPA}. The ions accelerated by \com{RTF-RPA} in the compression layer (S1) are trapped by \com{RTF-RPA} in the bulk and are further accelerated (S2). Ions initially at rest or accelerated by TNSA are not trapped and gain only little energy from the \com{RTF} passing them, (R), (T), respectively.}
    \label{fig:compression_mechanism}
\end{figure}

The reason for this large increase in ion energy lies in the higher terminal laser group velocity in less dense plasmas as before, but in this case there is no need for compensation by a longer pulse duration to facilitate more ion reflections. 
Instead, direct synchronization is achieved by the optimization of the plasma density profile. 
The ions can be synchronized at lower bulk density at a given short pulse duration, circumventing the limitation by dephasing due to pulse steepening at long pulse duration. 
Fig.~\ref{fig:3} shows the dynamics for the optimized two-layer \com{RTF-RPA} case. 
It can be seen that the front velocity reaches a maximum value exceeding $\beta_\mathrm{f}^\mathrm{PIC} \cong 0.77c$ (cf. $\beta_\mathrm{f} = 0.79$, Eqn.~\eqref{eqn:v_front}), which compares to only $\beta_\mathrm{f}^\mathrm{PIC}\cong 0.57c$ in the optimum homgeneous density case, cf Fig.~\ref{fig:1}. 
Again, panels b and c show the same characteristic dynamics as before: the most energetic ions comove with the point of laser reflection, which is located inside the plasma where the relativisticly corrected electron density surpasses the critical density and a charge separation is induced. 
Note that now that the laser penetrates the plasma much faster and hence the ion acceleration lasts much longer than before for the optimum case of a homogeneous plasma density in Fig.~\ref{fig:1}. 
While at 50 fs after the laser maximum arrival the ion energy is approximately equal to that of the optimum homogenous density case, in the following 100 fs the ions now continue to be accelerated up to 800 MeV. \\

What is left is a discussion how exactly the highest energy ions are injected and trapped in the fast moving charge separation field at \com{RTF}. 
The proton phase space is shown in Fig.~\ref{fig:compression_mechanism}a for several timesteps before and during the \com{RTF-RPA} process for the case of optimum ion acceleration with a front compression layer, cf.~Fig.~\ref{fig:3}. 
For comparison we also show the accelerating electric fields and electron densities in panels b and c, respectively. 
In panel c we scaled the density for each timestep individually for clarity, the thin dashed lines show the respective critical density $n_\mathrm{c}$ for reference. 
The ion acceleration starts early, already at $40\,$fs before the maximum arrival the ion gain energy at the front surface, while the electron density $n_\mathrm{e}$ is still seen to be quite similar to $n_\mathrm{e}/\gamma$. 
In this phase the laser pushes forward the surface while not yet being intense enough to accelerate electrons to relativistic energies. 
This changes at around $20\,$fs before the laser pulse maximum arrival. 
now the laser penetrates into the compression layer plasma by turning it relativisticly transparent at the surface. 
Up until between the laser maximum arrival and less than $10\,$fs later the most energetic ions can keep up with the \com{RTF}. 
At the same time, the compression layer rear surface expands into the less dense bulk similar to TNSA, also accelerating ions there. 
At that point, the compression layer has been turned fully transparent and the laser enters the less dense bulk. 
We can now observe three distinct ion acceleration groups, as the laser front sweeps through the bulk plasma. 
The first group, labeled \emph{R} in Fig.~\ref{fig:compression_mechanism} is that of bulk ions initially at rest. 
Those ions remain far too slow as they witness the charge separation field moving by. 
They are accelerated to only $\approx 0.1c$ while the front now passes by with $0.7$ to $0.8c$. 
The second group of ions is that pre-accelerated by TNSA at the compression layer rear surface (T). 
With their initial energy from TNSA they can comove somewhat longer with the laser front to reach $\approx. 0.2c$ before they are also outrun by the \com{RTF}. 
Only the ions that were pre-accelerated by \com{RTF-RPA} in the compression layer to $\approx 0.4 c$ already (S1) can be trapped in the \com{RTF} witnessed in panels b and c (S2). 
They undergo continued and repeated acceleration by \com{RTF-RPA} in the bulk and are thus accelerated to the final \com{RTF} velocity in the bulk and correspondingly high energies (beyond $800\MeV$ in this specific case). 
The shaded area indicates the large gain in energy by \com{RTF-RPA} for the optimum injected ions from the compression layer compared to the stationary bulk ions. 

\subsubsection{Lower laser Intensity}
\begin{figure}
    \centering
    \includegraphics[width=\linewidth]{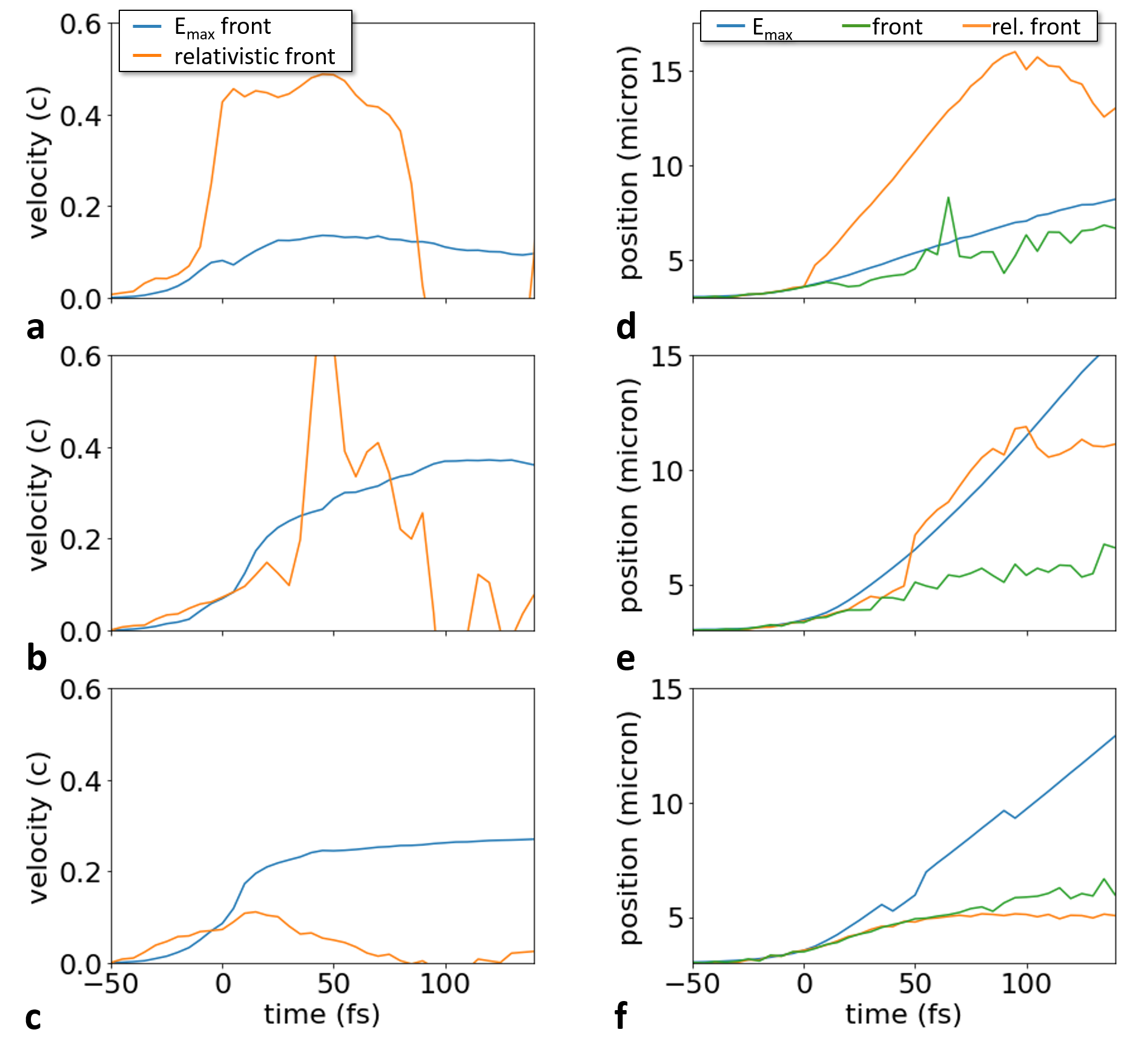}
    \caption{(a-c) Temporal evolution of the laser front velocity (orange) and velocity of the fastest ions(blue). (d-f) Temporal evolution of the respective positions   of the point of laser reflection (orange) for two close initial plasma densities and homogeneous profile ($n_1=0$) $n_0=4$ (a,d: transparent regime) and $n_0=5$ (c,f: hole boring regime) at $a_0=15$; and for the optimised \com{RTF-RPA} case with front compression, $n_0=2$, $n_1=7$, $d=1.6$ (b,e).
    }
    \label{fig:6}
\end{figure}
We now want to discuss an important benefit of the front compression layer specifically at smaller laser intensities, $a_0 \lesssim 20$, and short pulse durations, $t_\mathrm{FWHM} \lesssim 40\unit{fs}$, where \com{RTF-RPA} is less effective or even non-existent in homogeneous plasmas, cf. Fig.~\ref{fig:optT}. 
As reason for this problem we discussed that the matching between the \com{RTF} edge velocity and the ion acceleration can not be established in these cases. 
To illustrate this, we show two cases with homogeneous density $n_0$ close to each other (4 and 5 $n_c$) in Fig.~\ref{fig:6} tor $a_0=15$ and $t_\mathrm{FWHM}=40\unit{fs}$.
The problem can readily be seen in that for low density (a,d), the laser propagates too fast, while for a slightly higher density (c,f) the ions are reflected in hole boring like acceleration and the laser cannot keep up with them. 
Even as we tried densities closer to each other we could not find \com{RTF-RPA} in between, i.e. there appears to be a sudden jump from transparency regime to hole boring. \\
\begin{figure}
    \centering
    \includegraphics[width=\linewidth]{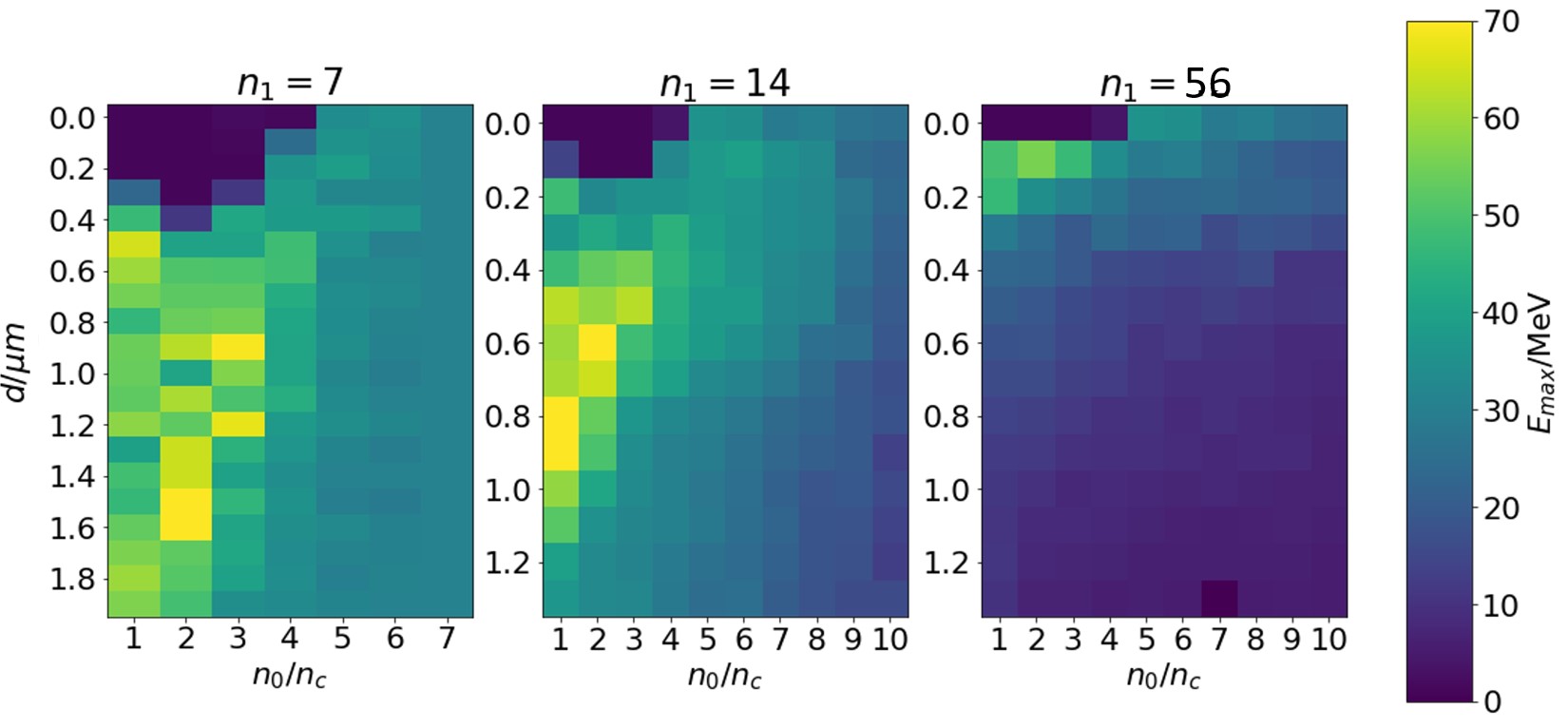}
    \caption{Maximum proton energy after the acceleration process with laser strength $a_0=15$, $140\unit{fs}$ after laser peak arrival on target as a function of rear layer density $n_0$ and front layer thickness $d$ for front layer density $n_1=7, 14, 56$ (from left to right).}
    \label{fig:7}
\end{figure}
Introducing a finite pre-compression layer changes the situation drastically, see Fig.~\ref{fig:7}. 
We can now see the same qualitative behaviour as for high laser intensity. 
The pre-compression layer shifts the optimum density of the bulk to lower values. 
At the same time the maximum ion energy increases with increasing thickness of the compression layer, reaching a maximum value of approximately twice that of the homogenous plasma. 
At this optimum \com{RTF-RPA} is well established, as indicated by the close vicinity of the most energetic ions and the point of the laser reflection witnessed in Fig.~\ref{fig:6}b,e and Fig.~\ref{fig:8}. 
The front compression layer keeps the laser slow at the beginning and gets faster after boring through, enabling the \com{RTF-RPA} process. 
Consequently, we can expect that for a pre-compressed target \com{RTF-RPA} can be effective even for low laser intensities, specifically in plasmas near the critical density and compression to modest density $7 \leq n_1 \leq 14$.
\begin{figure}
    \centering
    \includegraphics[width=\linewidth]{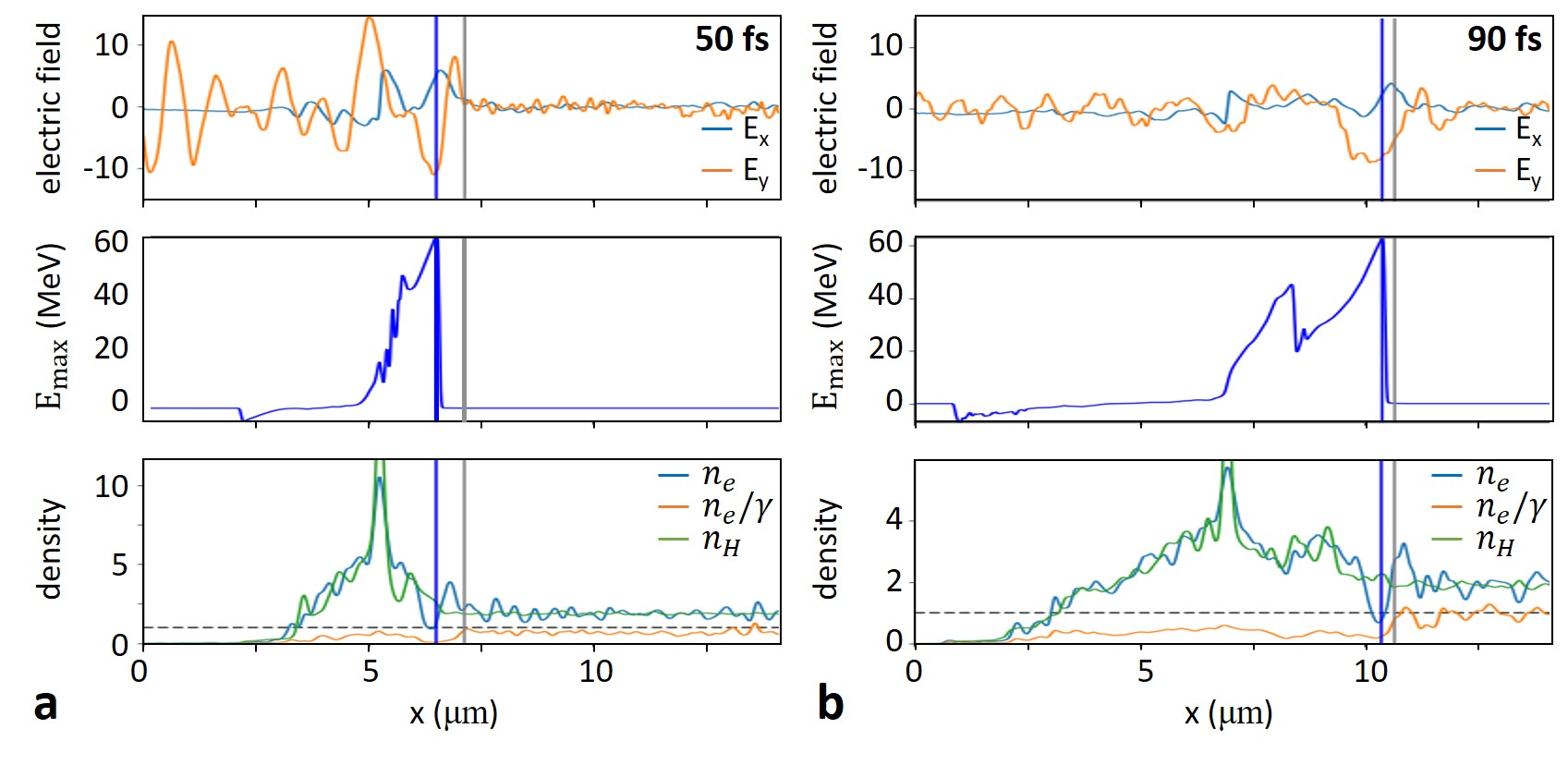}
    \caption{Plasma dynamics for the optimum front layer configuration at $a_0=15$: $n_0=2$, $n_1=7$, $d=1.6\unit{\mum}$. }
    \label{fig:8}
\end{figure}

\subsubsection{3D simulations}
\begin{figure}
    \centering
    \includegraphics[width=\linewidth]{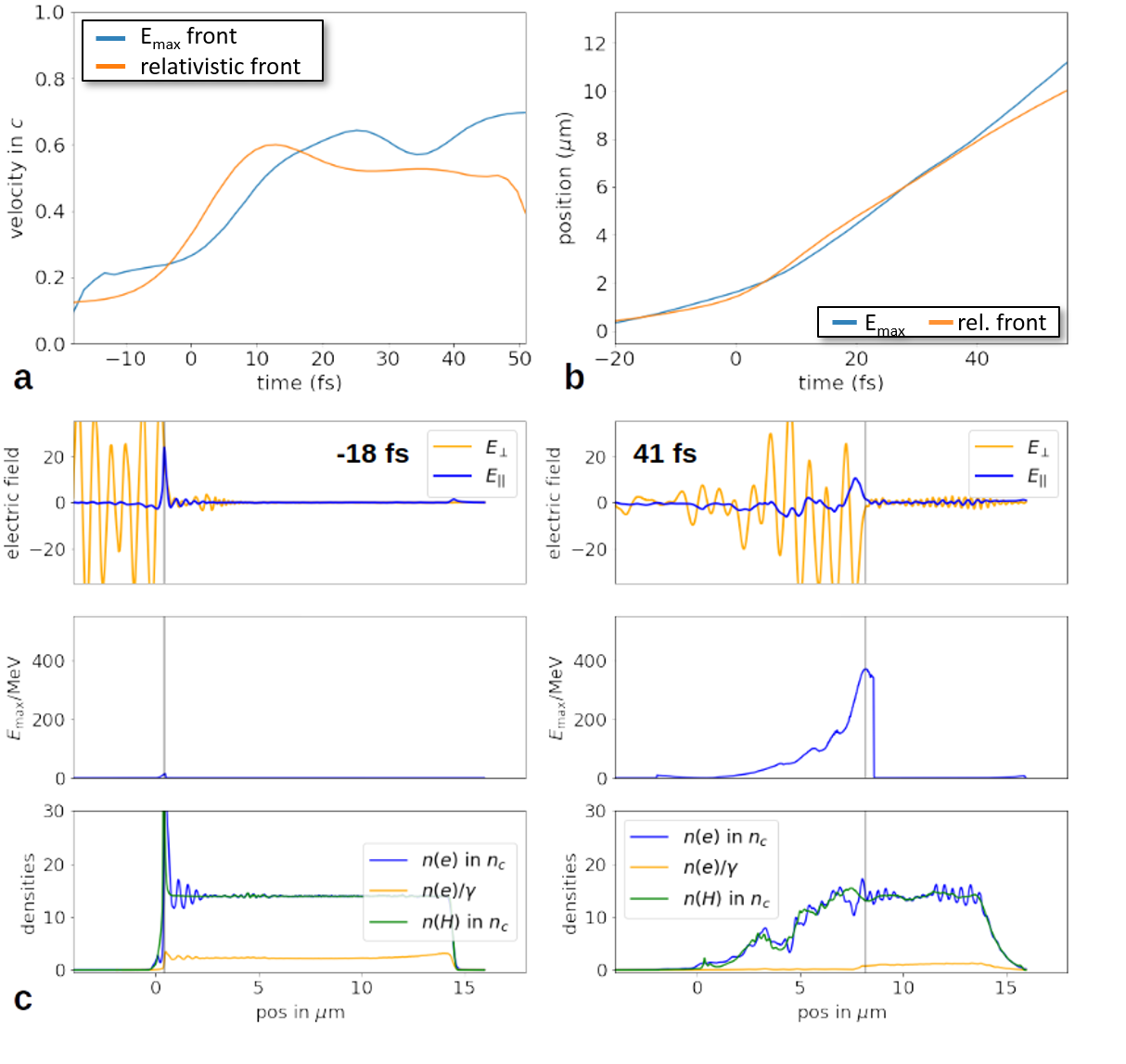}
    \caption{Lineouts along the laser axis showing the plasma dynamics for the 3D simulation. Target density is $n=13.9n_0$, which is the optimal case for a homogeneous target and the chosen laser of $a_0=50$ and a transversal width of $3.1\unit{\mum}$ at FWHM.}
    \label{fig:4}
\end{figure}
\begin{figure}
    \centering
    \includegraphics[width=\linewidth]{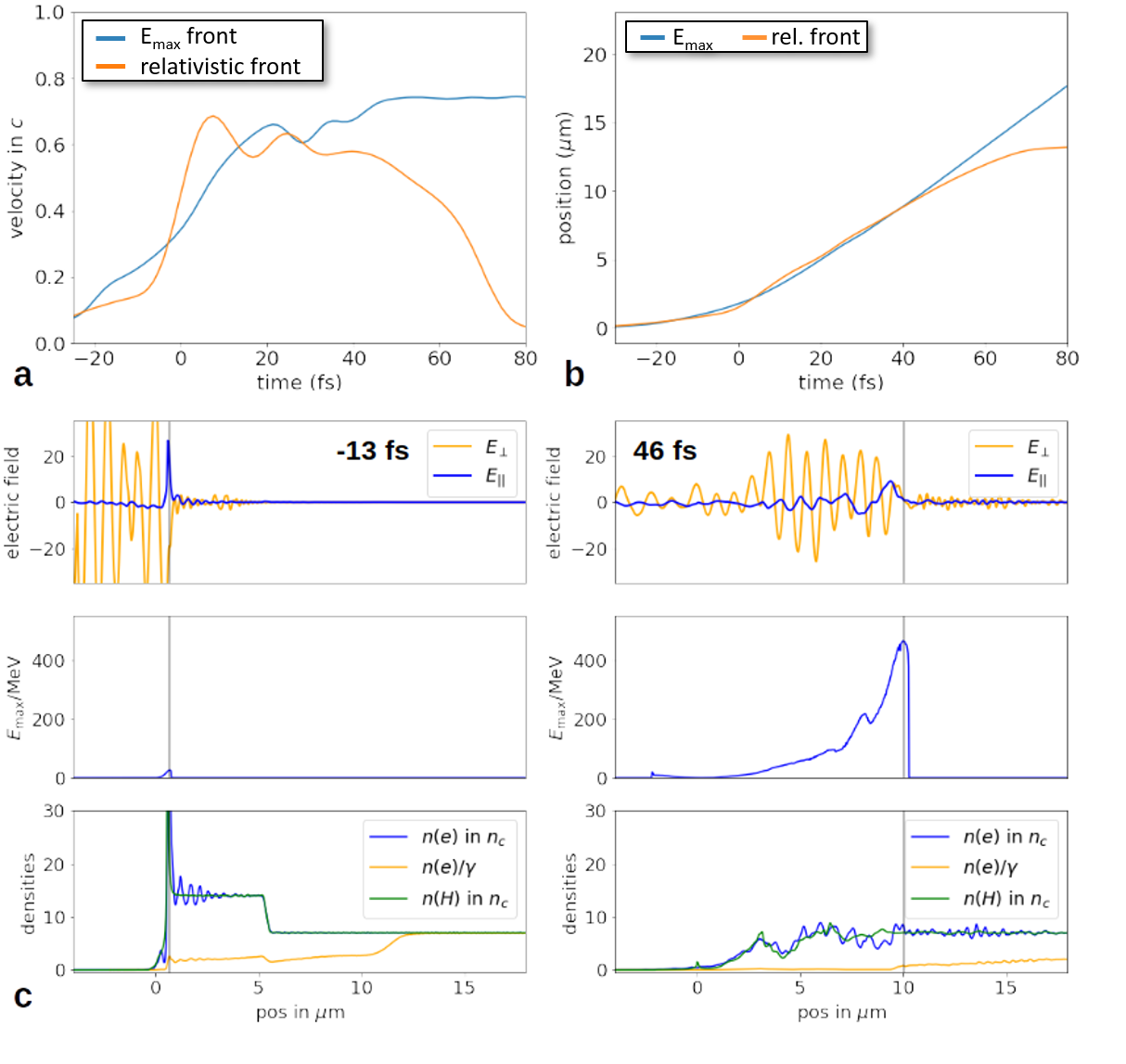}
    \caption{Lineouts along the laser axis showing the plasma dynamics for the 3D simulation with compressed front layer. An optimal configuration for the same laser parameters has a front layer density $n_\mathrm{f}=19n_0$, $d=5.4\unit{\mum}$ and a bulk density of $n_\mathrm{b}=7$. }
    \label{fig:5}
\end{figure}

It is important to note that the \com{RTF-RPA} process, and also the enhanced two-phase acceleration, can also be observed in 3D. 
This is to be expected since the \com{RTF-RPA} process is essentially a 1D process, but for future experimental realizations the applicability in 3D has to be checked and analyzed.  
We verified the effectiveness of the front compression layer by conducting exemplary 3D simulations for homogeneous and front compressed plasmas using PIConGPU~\cite{Bussmann2013}. 
The laser peak intensity and duration were kept as before in 1D, now with a tight focus of $3.1\unit{\mum}$ FWHM.
Figures~\ref{fig:4} and~\ref{fig:5} show the \com{RTF} edge and ion front velocities along with the profiles of electric fields, ion energies and plasma densities for the two optimal cases of a homogeneous and two-layered target, respectively. 
The ion energies are somewhat smaller in both cases than those observed in 1D geometry before, but again, we observe a substantial increase in the compression front plasma  
compared to the homogeneous plasma 
by more than $100\unit{MeV}$. \\ 
Thus, the qualitative behaviour of the system with varying parameters does not essentially differ, though some quantitative differences can be noted: the thickness of the compression layer that enhances the final energies is higher than for the 1D optimum, and also the density of the \com{RTF-RPA} optimum with a homogenous target is higher than in the 1D case. The latter is due to an increased laser strength due to self-focussing in the target, and transversally pushing the density away from the laser axis.

\section{Conclusions}
Synchronized ion acceleration by slow light in a near critical plasma is an effective means of dramatically increasing ion cutoff energies in comparison to the classical TNSA approach. 
We have shown that the the maximum \com{RTF-RPA} ion energies can be obtained from a simple semi-analytic model and the EOM was constructed. 
This allows to predict the temporal evolution during \com{RTF-RPA} as well as the optimum combination between the laser parameters pulse duration, intensity, and the plasma density.
We validated the model with a series of 1D PIC simulations, finding that the relationship between optimum density and pulse duration leads to maximum ion energy at the smallest density and highest possible pulse duration.
For fixed laser pulse energy the shortest pulse duration is expected to yield the most energetic ions at the respective optimum plasma density, as long as \com{RTF-RPA} is still active. 
For longer pulse durations we see a significant reduction of the simulated optimal energies compared to the theoretical maxima due to dephasing of the process for long interaction times. 
This in practice limits the pulse durations yielding optimized \com{RTF-RPA} to values below about 100 fs for current state-of-the art laser systems. \\
Even higher ion cutoff energies can be reached with an optimized plasma density profile. 
We demonstrated this in a proof-of-principal type geometry of a two-layer plasma density profile. 
By increasing the density in the front of the target and decreasing the density of the bulk in the rear, the \com{RTF} is synchronized better to the accelerating protons, which then can keep up with the former longer. 
We verified the principal effect in 3D simulations and at low laser intensity where \com{RTF-RPA} is not effective in homogeneous density plasmas. 
Those results could be extended in future by more simulations to find the optimum combination of plasma density and pulse shape for realistic conditions. 
This however requires the optimization of many parameters such as laser contrast, pulse duration, waist and dispersion, plasma shape at the same time. 
Here, the extension of the model to include e.g. 3D effects and the plasma density shaping by the laser could be essential accelerate such optimization. 

\section*{Data and code availability}
The code SMILEI~\cite{Derouillat2018} used for the simulations was published under the CeCILL license and PIConGPU was published under GNU General Public License 3~\cite{Bussmann2013}; input files and an exemplary PYTHON implementation of the numeric integration of the analytic model used for this publication is published under GNU Lesser General Public License 4.0~\cite{goethel2021scripts}. 

\section*{acknowledgments}
This work was partially  funded by the Center of Advanced Systems Understanding (CASUS), which is financed by Germany’s Federal Ministry of Education and Research (BMBF) and by the Saxon Ministry for Science, Culture and Tourism (SMWKT) with tax funds on the basis of the budget approved by the Saxon State Parliament.\\

\section*{Appendix}
The mathematical derivation of the factor $P$  which we introduced in $b$ in Eqn.~\eqref{eqn:v_front_K} \com{and} which changes the penetration velocity for linear polarization can be seen as follows, following the steps outlined in~\cite{Liu2020}. 
The derivation starts with Eqn. (2) in \cite{Liu2020} which reads for linear polarization (in our dimensionless unit system)
\begin{equation}
    p^{lin}_\perp = a_0 \left(\cos{\phi_0} -  \cos{r \phi_0}\right)
\end{equation}
where $r=\left(1-\beta_f\right)/\left(1+\beta_f\right)$. 
Hence the assumptions that lead to Eqn. (3) and (4) in  \cite{Liu2020} are still valid. 
In the following we use Eqn. (5) in \cite{Liu2020} in a slightly modified form, 
\begin{equation}
    E_{max}\cong n_e d / 2^{3/2},
    \label{eqn:modified}
\end{equation}
i.e. adding a factor $2^{-1/2}$. 
In our simulations for linear polarization this modified expression shows good agreement with all our simulations: E.g. in Fig.~\ref{fig:1} (c) we can extract from the blue line in the lower panel (i.e. $n_e(x)$) $d\approx 0.94\unit{\mum}\cdot 2\pi/\lambda=2.35\pi$ and with $n_{e,0}=8$ Eqn.~\eqref{eqn:modified} gives $E_{z,max}\cong 8\cdot 2.35\cdot \pi/2^{3/2}=20.7$, which compares reasonably well to $E_{z,max}\cong 20$ in the simulation, cp. blue line for $E_y$ in Fig.~\ref{fig:1}~(a) top panel. 
The difference of $1/2^{1/2}$ compared to Eqn. 5 in \cite{Liu2020} is clearly due to incomplete evacuation of the electron density in the plasma wave\com{, finite electron sheath thickness, and inhomogeneous background ion distribution}. \\
In fact, also for circular polarization this modified expression seems to fit reasonably well, cp. simulation Fig. 1~(a) in~\cite{Liu2020} $E_{z,max}\approx 25$ and the result of Eqn.~\eqref{eqn:modified} for the respective parameters ($n_e=0.8$, $\lambda = 1\unit{\mum}$, $d=15\unit{\mum}\cdot 2\pi/\lambda$) gives $E_{z,max}\cong 0.8\cdot 94.2/2^{3/2}=27$. 

We now follow the derivation of Eqn. (8) in \cite{Liu2020}. 
There, $\langle 1/\gamma\rangle_d$ was put into Eqn. (7) to account for the relativistic mass increase in the plasma wave. 
In the simulation of Fig. 1 in \cite{Liu2020}, using circular polarization, the authors state in that empirically "$\langle 1/\gamma\rangle_d\approx 0.006$ is about half of $1/2a_0$"$=1/4a_0$. 
Our simulation for $a_0=50$ at linear polarization shows that the average of the reciprocal of $\gamma$, $\langle1/\gamma\rangle_{d}\cong 0.018$, i.e. about $\langle1/\gamma\rangle_{d}\cong 1/a_0$. 
These two expressions put into Eqn. (7) in \cite{Liu2020}, together with the additional factor $1/2^{1/2}$ in Eqn.~\eqref{eqn:modified}, lead to Eqn. (8) in \cite{Liu2020} then reading for circular polarization
\begin{equation}
    E_{max} = \frac{\pi}{2^{1.5}} \left(1+\beta_f\right)\sqrt{4 a_0 n_0}. 
\end{equation}
and for linear polarization
\begin{equation}
    E_{max} = \frac{\pi}{2^{1.5}} \left(1+\beta_f\right)\sqrt{a_0 n_0}. 
\end{equation}
Together with Eqn. 4 we then recover Eqn. 9 in \cite{Liu2020} for circular polarization
\begin{equation}
    \frac{1-\beta_f}{(1+\beta_f)^2} = \frac{\pi^2}{8}\frac{n_0}{a_0}, 
\end{equation}
and obtain for linear polarization
\begin{equation}
    \frac{1-\beta_f}{(1+\beta_f)^2} = \frac{\pi^2}{8}\frac{n_0}{2 a_0}. 
\end{equation}
Hence,
\begin{equation}
    \frac{1-\beta_f}{(1+\beta_f)^2} = \frac{\pi^2}{16}\frac{P n_0}{a_0}. 
\end{equation}
This can be solved as outlined in \cite{Liu2020} for $\beta_f$, yielding Eqn.~\eqref{eqn:v_front_K}. It is in very good agreement with our 1D simulations both in terms of $\beta_f$ as well as $E_{z,max}$ and $\langle 1/\gamma\rangle_d$. 
%

\begin{thebibliography}{45}%
\makeatletter
\providecommand \@ifxundefined [1]{%
 \@ifx{#1\undefined}
}%
\providecommand \@ifnum [1]{%
 \ifnum #1\expandafter \@firstoftwo
 \else \expandafter \@secondoftwo
 \fi
}%
\providecommand \@ifx [1]{%
 \ifx #1\expandafter \@firstoftwo
 \else \expandafter \@secondoftwo
 \fi
}%
\providecommand \natexlab [1]{#1}%
\providecommand \enquote  [1]{``#1''}%
\providecommand \bibnamefont  [1]{#1}%
\providecommand \bibfnamefont [1]{#1}%
\providecommand \citenamefont [1]{#1}%
\providecommand \href@noop [0]{\@secondoftwo}%
\providecommand \href [0]{\begingroup \@sanitize@url \@href}%
\providecommand \@href[1]{\@@startlink{#1}\@@href}%
\providecommand \@@href[1]{\endgroup#1\@@endlink}%
\providecommand \@sanitize@url [0]{\catcode `\\12\catcode `\$12\catcode
  `\&12\catcode `\#12\catcode `\^12\catcode `\_12\catcode `\%12\relax}%
\providecommand \@@startlink[1]{}%
\providecommand \@@endlink[0]{}%
\providecommand \url  [0]{\begingroup\@sanitize@url \@url }%
\providecommand \@url [1]{\endgroup\@href {#1}{\urlprefix }}%
\providecommand \urlprefix  [0]{URL }%
\providecommand \Eprint [0]{\href }%
\providecommand \doibase [0]{http://dx.doi.org/}%
\providecommand \selectlanguage [0]{\@gobble}%
\providecommand \bibinfo  [0]{\@secondoftwo}%
\providecommand \bibfield  [0]{\@secondoftwo}%
\providecommand \translation [1]{[#1]}%
\providecommand \BibitemOpen [0]{}%
\providecommand \bibitemStop [0]{}%
\providecommand \bibitemNoStop [0]{.\EOS\space}%
\providecommand \EOS [0]{\spacefactor3000\relax}%
\providecommand \BibitemShut  [1]{\csname bibitem#1\endcsname}%
\let\auto@bib@innerbib\@empty
\bibitem [{\citenamefont {Bulanov}\ and\ \citenamefont
  {Khoroshkov}(2002)}]{Bulanov2002}%
  \BibitemOpen
  \bibfield  {author} {\bibinfo {author} {\bibfnamefont {S.~V.}\ \bibnamefont
  {Bulanov}}\ and\ \bibinfo {author} {\bibfnamefont {V.~S.}\ \bibnamefont
  {Khoroshkov}},\ }\href {\doibase 10.1134/1.1478534} {\bibfield  {journal}
  {\bibinfo  {journal} {Plasma Physics Reports}\ }\textbf {\bibinfo {volume}
  {28}},\ \bibinfo {pages} {453} (\bibinfo {year} {2002})}\BibitemShut
  {NoStop}%
\bibitem [{\citenamefont {Roth}\ \emph {et~al.}(2001)\citenamefont {Roth},
  \citenamefont {Cowan}, \citenamefont {Key}, \citenamefont {Hatchett},
  \citenamefont {Brown}, \citenamefont {Fountain}, \citenamefont {Johnson},
  \citenamefont {Pennington}, \citenamefont {Snavely}, \citenamefont {Wilks},
  \citenamefont {Yasuike}, \citenamefont {Ruhl}, \citenamefont {Pegoraro},
  \citenamefont {Bulanov}, \citenamefont {Campbell}, \citenamefont {Perry},\
  and\ \citenamefont {Powell}}]{Roth2001}%
  \BibitemOpen
  \bibfield  {author} {\bibinfo {author} {\bibfnamefont {M.}~\bibnamefont
  {Roth}}, \bibinfo {author} {\bibfnamefont {T.~E.}\ \bibnamefont {Cowan}},
  \bibinfo {author} {\bibfnamefont {M.~H.}\ \bibnamefont {Key}}, \bibinfo
  {author} {\bibfnamefont {S.~P.}\ \bibnamefont {Hatchett}}, \bibinfo {author}
  {\bibfnamefont {C.}~\bibnamefont {Brown}}, \bibinfo {author} {\bibfnamefont
  {W.}~\bibnamefont {Fountain}}, \bibinfo {author} {\bibfnamefont
  {J.}~\bibnamefont {Johnson}}, \bibinfo {author} {\bibfnamefont {D.~M.}\
  \bibnamefont {Pennington}}, \bibinfo {author} {\bibfnamefont {R.~A.}\
  \bibnamefont {Snavely}}, \bibinfo {author} {\bibfnamefont {S.~C.}\
  \bibnamefont {Wilks}}, \bibinfo {author} {\bibfnamefont {K.}~\bibnamefont
  {Yasuike}}, \bibinfo {author} {\bibfnamefont {H.}~\bibnamefont {Ruhl}},
  \bibinfo {author} {\bibfnamefont {F.}~\bibnamefont {Pegoraro}}, \bibinfo
  {author} {\bibfnamefont {S.~V.}\ \bibnamefont {Bulanov}}, \bibinfo {author}
  {\bibfnamefont {E.~M.}\ \bibnamefont {Campbell}}, \bibinfo {author}
  {\bibfnamefont {M.~D.}\ \bibnamefont {Perry}}, \ and\ \bibinfo {author}
  {\bibfnamefont {H.}~\bibnamefont {Powell}},\ }\href {\doibase
  10.1103/PhysRevLett.86.436} {\bibfield  {journal} {\bibinfo  {journal}
  {Physical Review Letters}\ }\textbf {\bibinfo {volume} {86}},\ \bibinfo
  {pages} {436} (\bibinfo {year} {2001})}\BibitemShut {NoStop}%
\bibitem [{\citenamefont {Fortney}\ \emph {et~al.}(2009)\citenamefont
  {Fortney}, \citenamefont {Glenzer}, \citenamefont {Koenig}, \citenamefont
  {Militzer}, \citenamefont {Saumon},\ and\ \citenamefont
  {Valencia}}]{Fortney2009}%
  \BibitemOpen
  \bibfield  {author} {\bibinfo {author} {\bibfnamefont {J.~J.}\ \bibnamefont
  {Fortney}}, \bibinfo {author} {\bibfnamefont {S.~H.}\ \bibnamefont
  {Glenzer}}, \bibinfo {author} {\bibfnamefont {M.}~\bibnamefont {Koenig}},
  \bibinfo {author} {\bibfnamefont {B.}~\bibnamefont {Militzer}}, \bibinfo
  {author} {\bibfnamefont {D.}~\bibnamefont {Saumon}}, \ and\ \bibinfo {author}
  {\bibfnamefont {D.}~\bibnamefont {Valencia}},\ }\href {\doibase
  10.1063/1.3101818} {\bibfield  {journal} {\bibinfo  {journal} {Physics of
  Plasmas}\ }\textbf {\bibinfo {volume} {16}},\ \bibinfo {pages} {041003}
  (\bibinfo {year} {2009})}\BibitemShut {NoStop}%
\bibitem [{\citenamefont {Schramm}\ \emph {et~al.}(2017)\citenamefont
  {Schramm}, \citenamefont {Bussmann}, \citenamefont {Irman}, \citenamefont
  {Siebold}, \citenamefont {Zeil}, \citenamefont {Albach}, \citenamefont
  {Bernert}, \citenamefont {Bock}, \citenamefont {Brack}, \citenamefont
  {Branco}, \citenamefont {Couperus}, \citenamefont {Cowan}, \citenamefont
  {Debus}, \citenamefont {Eisenmann}, \citenamefont {Garten}, \citenamefont
  {Gebhardt}, \citenamefont {Grams}, \citenamefont {Helbig}, \citenamefont
  {Huebl}, \citenamefont {Kluge}, \citenamefont {K{\"{o}}hler}, \citenamefont
  {Kr{\"{a}}mer}, \citenamefont {Kraft}, \citenamefont {Kroll}, \citenamefont
  {Kuntzsch}, \citenamefont {Lehnert}, \citenamefont {Loeser}, \citenamefont
  {Metzkes}, \citenamefont {Michel}, \citenamefont {Obst}, \citenamefont
  {Pausch}, \citenamefont {Rehwald}, \citenamefont {Sauerbrey}, \citenamefont
  {Schlenvoigt}, \citenamefont {Steiniger},\ and\ \citenamefont
  {Zarini}}]{Schramm2017}%
  \BibitemOpen
  \bibfield  {author} {\bibinfo {author} {\bibfnamefont {U.}~\bibnamefont
  {Schramm}}, \bibinfo {author} {\bibfnamefont {M.}~\bibnamefont {Bussmann}},
  \bibinfo {author} {\bibfnamefont {A.}~\bibnamefont {Irman}}, \bibinfo
  {author} {\bibfnamefont {M.}~\bibnamefont {Siebold}}, \bibinfo {author}
  {\bibfnamefont {K.}~\bibnamefont {Zeil}}, \bibinfo {author} {\bibfnamefont
  {D.}~\bibnamefont {Albach}}, \bibinfo {author} {\bibfnamefont
  {C.}~\bibnamefont {Bernert}}, \bibinfo {author} {\bibfnamefont
  {S.}~\bibnamefont {Bock}}, \bibinfo {author} {\bibfnamefont {F.}~\bibnamefont
  {Brack}}, \bibinfo {author} {\bibfnamefont {J.}~\bibnamefont {Branco}},
  \bibinfo {author} {\bibfnamefont {J.~P.}\ \bibnamefont {Couperus}}, \bibinfo
  {author} {\bibfnamefont {T.~E.}\ \bibnamefont {Cowan}}, \bibinfo {author}
  {\bibfnamefont {A.}~\bibnamefont {Debus}}, \bibinfo {author} {\bibfnamefont
  {C.}~\bibnamefont {Eisenmann}}, \bibinfo {author} {\bibfnamefont
  {M.}~\bibnamefont {Garten}}, \bibinfo {author} {\bibfnamefont
  {R.}~\bibnamefont {Gebhardt}}, \bibinfo {author} {\bibfnamefont
  {S.}~\bibnamefont {Grams}}, \bibinfo {author} {\bibfnamefont
  {U.}~\bibnamefont {Helbig}}, \bibinfo {author} {\bibfnamefont
  {A.}~\bibnamefont {Huebl}}, \bibinfo {author} {\bibfnamefont
  {T.}~\bibnamefont {Kluge}}, \bibinfo {author} {\bibfnamefont
  {A.}~\bibnamefont {K{\"{o}}hler}}, \bibinfo {author} {\bibfnamefont {J.~M.}\
  \bibnamefont {Kr{\"{a}}mer}}, \bibinfo {author} {\bibfnamefont
  {S.}~\bibnamefont {Kraft}}, \bibinfo {author} {\bibfnamefont
  {F.}~\bibnamefont {Kroll}}, \bibinfo {author} {\bibfnamefont
  {M.}~\bibnamefont {Kuntzsch}}, \bibinfo {author} {\bibfnamefont
  {U.}~\bibnamefont {Lehnert}}, \bibinfo {author} {\bibfnamefont
  {M.}~\bibnamefont {Loeser}}, \bibinfo {author} {\bibfnamefont
  {J.}~\bibnamefont {Metzkes}}, \bibinfo {author} {\bibfnamefont
  {P.}~\bibnamefont {Michel}}, \bibinfo {author} {\bibfnamefont
  {L.}~\bibnamefont {Obst}}, \bibinfo {author} {\bibfnamefont {R.}~\bibnamefont
  {Pausch}}, \bibinfo {author} {\bibfnamefont {M.}~\bibnamefont {Rehwald}},
  \bibinfo {author} {\bibfnamefont {R.}~\bibnamefont {Sauerbrey}}, \bibinfo
  {author} {\bibfnamefont {H.~P.}\ \bibnamefont {Schlenvoigt}}, \bibinfo
  {author} {\bibfnamefont {K.}~\bibnamefont {Steiniger}}, \ and\ \bibinfo
  {author} {\bibfnamefont {O.}~\bibnamefont {Zarini}},\ }\href {\doibase
  10.1088/1742-6596/874/1/012028} {\bibfield  {journal} {\bibinfo  {journal}
  {Journal of Physics: Conference Series}\ }\textbf {\bibinfo {volume} {874}},\
  \bibinfo {pages} {012028} (\bibinfo {year} {2017})}\BibitemShut {NoStop}%
\bibitem [{\citenamefont {Danson}\ \emph {et~al.}(2019)\citenamefont {Danson},
  \citenamefont {Haefner}, \citenamefont {Bromage}, \citenamefont {Butcher},
  \citenamefont {Chanteloup}, \citenamefont {Chowdhury}, \citenamefont
  {Galvanauskas}, \citenamefont {Gizzi}, \citenamefont {Hein}, \citenamefont
  {Hillier}, \citenamefont {Hopps}, \citenamefont {Kato}, \citenamefont
  {Khazanov}, \citenamefont {Kodama}, \citenamefont {Korn}, \citenamefont {Li},
  \citenamefont {Li}, \citenamefont {Limpert}, \citenamefont {Ma},
  \citenamefont {Nam}, \citenamefont {Neely}, \citenamefont {Papadopoulos},
  \citenamefont {Penman}, \citenamefont {Qian}, \citenamefont {Rocca},
  \citenamefont {Shaykin}, \citenamefont {Siders}, \citenamefont {Spindloe},
  \citenamefont {Szatm{\'{a}}ri}, \citenamefont {Trines}, \citenamefont {Zhu},
  \citenamefont {Zhu},\ and\ \citenamefont {Zuegel}}]{Danson2019}%
  \BibitemOpen
  \bibfield  {author} {\bibinfo {author} {\bibfnamefont {C.~N.}\ \bibnamefont
  {Danson}}, \bibinfo {author} {\bibfnamefont {C.}~\bibnamefont {Haefner}},
  \bibinfo {author} {\bibfnamefont {J.}~\bibnamefont {Bromage}}, \bibinfo
  {author} {\bibfnamefont {T.}~\bibnamefont {Butcher}}, \bibinfo {author}
  {\bibfnamefont {J.~C.~F.}\ \bibnamefont {Chanteloup}}, \bibinfo {author}
  {\bibfnamefont {E.~A.}\ \bibnamefont {Chowdhury}}, \bibinfo {author}
  {\bibfnamefont {A.}~\bibnamefont {Galvanauskas}}, \bibinfo {author}
  {\bibfnamefont {L.~A.}\ \bibnamefont {Gizzi}}, \bibinfo {author}
  {\bibfnamefont {J.}~\bibnamefont {Hein}}, \bibinfo {author} {\bibfnamefont
  {D.~I.}\ \bibnamefont {Hillier}}, \bibinfo {author} {\bibfnamefont {N.~W.}\
  \bibnamefont {Hopps}}, \bibinfo {author} {\bibfnamefont {Y.}~\bibnamefont
  {Kato}}, \bibinfo {author} {\bibfnamefont {E.~A.}\ \bibnamefont {Khazanov}},
  \bibinfo {author} {\bibfnamefont {R.}~\bibnamefont {Kodama}}, \bibinfo
  {author} {\bibfnamefont {G.}~\bibnamefont {Korn}}, \bibinfo {author}
  {\bibfnamefont {R.}~\bibnamefont {Li}}, \bibinfo {author} {\bibfnamefont
  {Y.}~\bibnamefont {Li}}, \bibinfo {author} {\bibfnamefont {J.}~\bibnamefont
  {Limpert}}, \bibinfo {author} {\bibfnamefont {J.}~\bibnamefont {Ma}},
  \bibinfo {author} {\bibfnamefont {C.~H.}\ \bibnamefont {Nam}}, \bibinfo
  {author} {\bibfnamefont {D.}~\bibnamefont {Neely}}, \bibinfo {author}
  {\bibfnamefont {D.}~\bibnamefont {Papadopoulos}}, \bibinfo {author}
  {\bibfnamefont {R.~R.}\ \bibnamefont {Penman}}, \bibinfo {author}
  {\bibfnamefont {L.}~\bibnamefont {Qian}}, \bibinfo {author} {\bibfnamefont
  {J.~J.}\ \bibnamefont {Rocca}}, \bibinfo {author} {\bibfnamefont {A.~A.}\
  \bibnamefont {Shaykin}}, \bibinfo {author} {\bibfnamefont {C.~W.}\
  \bibnamefont {Siders}}, \bibinfo {author} {\bibfnamefont {C.}~\bibnamefont
  {Spindloe}}, \bibinfo {author} {\bibfnamefont {S.}~\bibnamefont
  {Szatm{\'{a}}ri}}, \bibinfo {author} {\bibfnamefont {R.~M.}\ \bibnamefont
  {Trines}}, \bibinfo {author} {\bibfnamefont {J.}~\bibnamefont {Zhu}},
  \bibinfo {author} {\bibfnamefont {P.}~\bibnamefont {Zhu}}, \ and\ \bibinfo
  {author} {\bibfnamefont {J.~D.}\ \bibnamefont {Zuegel}},\ }\href {\doibase
  10.1017/hpl.2019.36} {\bibfield  {journal} {\bibinfo  {journal} {High Power
  Laser Science and Engineering}\ }\textbf {\bibinfo {volume} {7}} (\bibinfo
  {year} {2019}),\ 10.1017/hpl.2019.36}\BibitemShut {NoStop}%
\bibitem [{\citenamefont {Obst-Huebl}\ \emph {et~al.}(2018)\citenamefont
  {Obst-Huebl}, \citenamefont {Ziegler}, \citenamefont {Brack}, \citenamefont
  {Branco}, \citenamefont {Bussmann}, \citenamefont {Cowan}, \citenamefont
  {Curry}, \citenamefont {Fiuza}, \citenamefont {Garten}, \citenamefont
  {Gauthier}, \citenamefont {G{\"{o}}de}, \citenamefont {Glenzer},
  \citenamefont {Huebl}, \citenamefont {Irman}, \citenamefont {Kim},
  \citenamefont {Kluge}, \citenamefont {Kraft}, \citenamefont {Kroll},
  \citenamefont {Metzkes-Ng}, \citenamefont {Pausch}, \citenamefont {Prencipe},
  \citenamefont {Rehwald}, \citenamefont {Roedel}, \citenamefont {Schlenvoigt},
  \citenamefont {Schramm},\ and\ \citenamefont {Zeil}}]{Obst-Huebl2018}%
  \BibitemOpen
  \bibfield  {author} {\bibinfo {author} {\bibfnamefont {L.}~\bibnamefont
  {Obst-Huebl}}, \bibinfo {author} {\bibfnamefont {T.}~\bibnamefont {Ziegler}},
  \bibinfo {author} {\bibfnamefont {F.~E.}\ \bibnamefont {Brack}}, \bibinfo
  {author} {\bibfnamefont {J.}~\bibnamefont {Branco}}, \bibinfo {author}
  {\bibfnamefont {M.}~\bibnamefont {Bussmann}}, \bibinfo {author}
  {\bibfnamefont {T.~E.}\ \bibnamefont {Cowan}}, \bibinfo {author}
  {\bibfnamefont {C.~B.}\ \bibnamefont {Curry}}, \bibinfo {author}
  {\bibfnamefont {F.}~\bibnamefont {Fiuza}}, \bibinfo {author} {\bibfnamefont
  {M.}~\bibnamefont {Garten}}, \bibinfo {author} {\bibfnamefont
  {M.}~\bibnamefont {Gauthier}}, \bibinfo {author} {\bibfnamefont
  {S.}~\bibnamefont {G{\"{o}}de}}, \bibinfo {author} {\bibfnamefont {S.~H.}\
  \bibnamefont {Glenzer}}, \bibinfo {author} {\bibfnamefont {A.}~\bibnamefont
  {Huebl}}, \bibinfo {author} {\bibfnamefont {A.}~\bibnamefont {Irman}},
  \bibinfo {author} {\bibfnamefont {J.~B.}\ \bibnamefont {Kim}}, \bibinfo
  {author} {\bibfnamefont {T.}~\bibnamefont {Kluge}}, \bibinfo {author}
  {\bibfnamefont {S.~D.}\ \bibnamefont {Kraft}}, \bibinfo {author}
  {\bibfnamefont {F.}~\bibnamefont {Kroll}}, \bibinfo {author} {\bibfnamefont
  {J.}~\bibnamefont {Metzkes-Ng}}, \bibinfo {author} {\bibfnamefont
  {R.}~\bibnamefont {Pausch}}, \bibinfo {author} {\bibfnamefont
  {I.}~\bibnamefont {Prencipe}}, \bibinfo {author} {\bibfnamefont
  {M.}~\bibnamefont {Rehwald}}, \bibinfo {author} {\bibfnamefont
  {C.}~\bibnamefont {Roedel}}, \bibinfo {author} {\bibfnamefont {H.~P.}\
  \bibnamefont {Schlenvoigt}}, \bibinfo {author} {\bibfnamefont
  {U.}~\bibnamefont {Schramm}}, \ and\ \bibinfo {author} {\bibfnamefont
  {K.}~\bibnamefont {Zeil}},\ }\href {\doibase 10.1038/s41467-018-07756-z}
  {\bibfield  {journal} {\bibinfo  {journal} {Nature Communications}\ }\textbf
  {\bibinfo {volume} {9}},\ \bibinfo {pages} {5292} (\bibinfo {year}
  {2018})}\BibitemShut {NoStop}%
\bibitem [{\citenamefont {Brack}\ \emph {et~al.}(2020)\citenamefont {Brack},
  \citenamefont {Kroll}, \citenamefont {Gaus}, \citenamefont {Bernert},
  \citenamefont {Beyreuther}, \citenamefont {Cowan}, \citenamefont {Karsch},
  \citenamefont {Kraft}, \citenamefont {Kunz-Schughart}, \citenamefont
  {Lessmann}, \citenamefont {Metzkes-Ng}, \citenamefont {Obst-Huebl},
  \citenamefont {Pawelke}, \citenamefont {Rehwald}, \citenamefont
  {Schlenvoigt}, \citenamefont {Schramm}, \citenamefont {Sobiella},
  \citenamefont {Szab{\'{o}}}, \citenamefont {Ziegler},\ and\ \citenamefont
  {Zeil}}]{Brack2020}%
  \BibitemOpen
  \bibfield  {author} {\bibinfo {author} {\bibfnamefont {F.~E.}\ \bibnamefont
  {Brack}}, \bibinfo {author} {\bibfnamefont {F.}~\bibnamefont {Kroll}},
  \bibinfo {author} {\bibfnamefont {L.}~\bibnamefont {Gaus}}, \bibinfo {author}
  {\bibfnamefont {C.}~\bibnamefont {Bernert}}, \bibinfo {author} {\bibfnamefont
  {E.}~\bibnamefont {Beyreuther}}, \bibinfo {author} {\bibfnamefont {T.~E.}\
  \bibnamefont {Cowan}}, \bibinfo {author} {\bibfnamefont {L.}~\bibnamefont
  {Karsch}}, \bibinfo {author} {\bibfnamefont {S.}~\bibnamefont {Kraft}},
  \bibinfo {author} {\bibfnamefont {L.~A.}\ \bibnamefont {Kunz-Schughart}},
  \bibinfo {author} {\bibfnamefont {E.}~\bibnamefont {Lessmann}}, \bibinfo
  {author} {\bibfnamefont {J.}~\bibnamefont {Metzkes-Ng}}, \bibinfo {author}
  {\bibfnamefont {L.}~\bibnamefont {Obst-Huebl}}, \bibinfo {author}
  {\bibfnamefont {J.}~\bibnamefont {Pawelke}}, \bibinfo {author} {\bibfnamefont
  {M.}~\bibnamefont {Rehwald}}, \bibinfo {author} {\bibfnamefont {H.~P.}\
  \bibnamefont {Schlenvoigt}}, \bibinfo {author} {\bibfnamefont
  {U.}~\bibnamefont {Schramm}}, \bibinfo {author} {\bibfnamefont
  {M.}~\bibnamefont {Sobiella}}, \bibinfo {author} {\bibfnamefont {E.~R.}\
  \bibnamefont {Szab{\'{o}}}}, \bibinfo {author} {\bibfnamefont
  {T.}~\bibnamefont {Ziegler}}, \ and\ \bibinfo {author} {\bibfnamefont
  {K.}~\bibnamefont {Zeil}},\ }\href {\doibase 10.1038/s41598-020-65775-7}
  {\bibfield  {journal} {\bibinfo  {journal} {Scientific Reports}\ }\textbf
  {\bibinfo {volume} {10}},\ \bibinfo {pages} {1} (\bibinfo {year}
  {2020})}\BibitemShut {NoStop}%
\bibitem [{\citenamefont {Ziegler}\ \emph {et~al.}(2021)\citenamefont
  {Ziegler}, \citenamefont {Albach}, \citenamefont {Bernert}, \citenamefont
  {Bock}, \citenamefont {Brack}, \citenamefont {Cowan}, \citenamefont {Dover},
  \citenamefont {Garten}, \citenamefont {Gaus}, \citenamefont {Gebhardt},
  \citenamefont {Goethel}, \citenamefont {Helbig}, \citenamefont {Irman},
  \citenamefont {Kiriyama}, \citenamefont {Kluge}, \citenamefont {Kon},
  \citenamefont {Kraft}, \citenamefont {Kroll}, \citenamefont {Loeser},
  \citenamefont {Metzkes-Ng}, \citenamefont {Nishiuchi}, \citenamefont
  {Obst-Huebl}, \citenamefont {P{\"{u}}schel}, \citenamefont {Rehwald},
  \citenamefont {Schlenvoigt}, \citenamefont {Schramm},\ and\ \citenamefont
  {Zeil}}]{Ziegler2020}%
  \BibitemOpen
  \bibfield  {author} {\bibinfo {author} {\bibfnamefont {T.}~\bibnamefont
  {Ziegler}}, \bibinfo {author} {\bibfnamefont {D.}~\bibnamefont {Albach}},
  \bibinfo {author} {\bibfnamefont {C.}~\bibnamefont {Bernert}}, \bibinfo
  {author} {\bibfnamefont {S.}~\bibnamefont {Bock}}, \bibinfo {author}
  {\bibfnamefont {F.~E.}\ \bibnamefont {Brack}}, \bibinfo {author}
  {\bibfnamefont {T.~E.}\ \bibnamefont {Cowan}}, \bibinfo {author}
  {\bibfnamefont {N.~P.}\ \bibnamefont {Dover}}, \bibinfo {author}
  {\bibfnamefont {M.}~\bibnamefont {Garten}}, \bibinfo {author} {\bibfnamefont
  {L.}~\bibnamefont {Gaus}}, \bibinfo {author} {\bibfnamefont {R.}~\bibnamefont
  {Gebhardt}}, \bibinfo {author} {\bibfnamefont {I.}~\bibnamefont {Goethel}},
  \bibinfo {author} {\bibfnamefont {U.}~\bibnamefont {Helbig}}, \bibinfo
  {author} {\bibfnamefont {A.}~\bibnamefont {Irman}}, \bibinfo {author}
  {\bibfnamefont {H.}~\bibnamefont {Kiriyama}}, \bibinfo {author}
  {\bibfnamefont {T.}~\bibnamefont {Kluge}}, \bibinfo {author} {\bibfnamefont
  {A.}~\bibnamefont {Kon}}, \bibinfo {author} {\bibfnamefont {S.}~\bibnamefont
  {Kraft}}, \bibinfo {author} {\bibfnamefont {F.}~\bibnamefont {Kroll}},
  \bibinfo {author} {\bibfnamefont {M.}~\bibnamefont {Loeser}}, \bibinfo
  {author} {\bibfnamefont {J.}~\bibnamefont {Metzkes-Ng}}, \bibinfo {author}
  {\bibfnamefont {M.}~\bibnamefont {Nishiuchi}}, \bibinfo {author}
  {\bibfnamefont {L.}~\bibnamefont {Obst-Huebl}}, \bibinfo {author}
  {\bibfnamefont {T.}~\bibnamefont {P{\"{u}}schel}}, \bibinfo {author}
  {\bibfnamefont {M.}~\bibnamefont {Rehwald}}, \bibinfo {author} {\bibfnamefont
  {H.~P.}\ \bibnamefont {Schlenvoigt}}, \bibinfo {author} {\bibfnamefont
  {U.}~\bibnamefont {Schramm}}, \ and\ \bibinfo {author} {\bibfnamefont
  {K.}~\bibnamefont {Zeil}},\ }\href {\doibase 10.1038/s41598-021-86547-x}
  {\bibfield  {journal} {\bibinfo  {journal} {Scientific Reports}\ }\textbf
  {\bibinfo {volume} {11}},\ \bibinfo {pages} {7338} (\bibinfo {year}
  {2021})}\BibitemShut {NoStop}%
\bibitem [{\citenamefont {Huebl}\ \emph {et~al.}(2020)\citenamefont {Huebl},
  \citenamefont {Rehwald}, \citenamefont {Obst-Huebl}, \citenamefont {Ziegler},
  \citenamefont {Garten}, \citenamefont {Widera}, \citenamefont {Zeil},
  \citenamefont {Cowan}, \citenamefont {Bussmann}, \citenamefont {Schramm},\
  and\ \citenamefont {Kluge}}]{Huebl2020}%
  \BibitemOpen
  \bibfield  {author} {\bibinfo {author} {\bibfnamefont {A.}~\bibnamefont
  {Huebl}}, \bibinfo {author} {\bibfnamefont {M.}~\bibnamefont {Rehwald}},
  \bibinfo {author} {\bibfnamefont {L.}~\bibnamefont {Obst-Huebl}}, \bibinfo
  {author} {\bibfnamefont {T.}~\bibnamefont {Ziegler}}, \bibinfo {author}
  {\bibfnamefont {M.}~\bibnamefont {Garten}}, \bibinfo {author} {\bibfnamefont
  {R.}~\bibnamefont {Widera}}, \bibinfo {author} {\bibfnamefont
  {K.}~\bibnamefont {Zeil}}, \bibinfo {author} {\bibfnamefont {T.~E.}\
  \bibnamefont {Cowan}}, \bibinfo {author} {\bibfnamefont {M.}~\bibnamefont
  {Bussmann}}, \bibinfo {author} {\bibfnamefont {U.}~\bibnamefont {Schramm}}, \
  and\ \bibinfo {author} {\bibfnamefont {T.}~\bibnamefont {Kluge}},\ }\href
  {\doibase 10.1088/1361-6587/abbe33} {\bibfield  {journal} {\bibinfo
  {journal} {Plasma Physics and Controlled Fusion}\ }\textbf {\bibinfo {volume}
  {62}} (\bibinfo {year} {2020}),\ 10.1088/1361-6587/abbe33}\BibitemShut
  {NoStop}%
\bibitem [{\citenamefont {Gaillard}\ \emph {et~al.}(2011)\citenamefont
  {Gaillard}, \citenamefont {Kluge}, \citenamefont {Flippo}, \citenamefont
  {Bussmann}, \citenamefont {Gall}, \citenamefont {Lockard}, \citenamefont
  {Geissel}, \citenamefont {Offermann}, \citenamefont {Schollmeier},
  \citenamefont {Sentoku},\ and\ \citenamefont {Cowan}}]{Gaillard2011}%
  \BibitemOpen
  \bibfield  {author} {\bibinfo {author} {\bibfnamefont {S.~A.}\ \bibnamefont
  {Gaillard}}, \bibinfo {author} {\bibfnamefont {T.}~\bibnamefont {Kluge}},
  \bibinfo {author} {\bibfnamefont {K.~A.}\ \bibnamefont {Flippo}}, \bibinfo
  {author} {\bibfnamefont {M.}~\bibnamefont {Bussmann}}, \bibinfo {author}
  {\bibfnamefont {B.}~\bibnamefont {Gall}}, \bibinfo {author} {\bibfnamefont
  {T.}~\bibnamefont {Lockard}}, \bibinfo {author} {\bibfnamefont
  {M.}~\bibnamefont {Geissel}}, \bibinfo {author} {\bibfnamefont {D.~T.}\
  \bibnamefont {Offermann}}, \bibinfo {author} {\bibfnamefont {M.}~\bibnamefont
  {Schollmeier}}, \bibinfo {author} {\bibfnamefont {Y.}~\bibnamefont
  {Sentoku}}, \ and\ \bibinfo {author} {\bibfnamefont {T.~E.}\ \bibnamefont
  {Cowan}},\ }\href {\doibase 10.1063/1.3575624} {\bibfield  {journal}
  {\bibinfo  {journal} {Physics of Plasmas}\ }\textbf {\bibinfo {volume}
  {18}},\ \bibinfo {pages} {056710} (\bibinfo {year} {2011})}\BibitemShut
  {NoStop}%
\bibitem [{\citenamefont {Hilz}\ \emph {et~al.}(2018)\citenamefont {Hilz},
  \citenamefont {Ostermayr}, \citenamefont {Huebl}, \citenamefont {Bagnoud},
  \citenamefont {Borm}, \citenamefont {Bussmann}, \citenamefont {Gallei},
  \citenamefont {Gebhard}, \citenamefont {Haffa}, \citenamefont {Hartmann},
  \citenamefont {Kluge}, \citenamefont {Lindner}, \citenamefont {Neumayr},
  \citenamefont {Schaefer}, \citenamefont {Schramm}, \citenamefont {Thirolf},
  \citenamefont {R{\"{o}}sch}, \citenamefont {Wagner}, \citenamefont
  {Zielbauer},\ and\ \citenamefont {Schreiber}}]{Hilz2018}%
  \BibitemOpen
  \bibfield  {author} {\bibinfo {author} {\bibfnamefont {P.}~\bibnamefont
  {Hilz}}, \bibinfo {author} {\bibfnamefont {T.~M.}\ \bibnamefont {Ostermayr}},
  \bibinfo {author} {\bibfnamefont {A.}~\bibnamefont {Huebl}}, \bibinfo
  {author} {\bibfnamefont {V.}~\bibnamefont {Bagnoud}}, \bibinfo {author}
  {\bibfnamefont {B.}~\bibnamefont {Borm}}, \bibinfo {author} {\bibfnamefont
  {M.}~\bibnamefont {Bussmann}}, \bibinfo {author} {\bibfnamefont
  {M.}~\bibnamefont {Gallei}}, \bibinfo {author} {\bibfnamefont
  {J.}~\bibnamefont {Gebhard}}, \bibinfo {author} {\bibfnamefont
  {D.}~\bibnamefont {Haffa}}, \bibinfo {author} {\bibfnamefont
  {J.}~\bibnamefont {Hartmann}}, \bibinfo {author} {\bibfnamefont
  {T.}~\bibnamefont {Kluge}}, \bibinfo {author} {\bibfnamefont {F.~H.}\
  \bibnamefont {Lindner}}, \bibinfo {author} {\bibfnamefont {P.}~\bibnamefont
  {Neumayr}}, \bibinfo {author} {\bibfnamefont {C.~G.}\ \bibnamefont
  {Schaefer}}, \bibinfo {author} {\bibfnamefont {U.}~\bibnamefont {Schramm}},
  \bibinfo {author} {\bibfnamefont {P.~G.}\ \bibnamefont {Thirolf}}, \bibinfo
  {author} {\bibfnamefont {T.~F.}\ \bibnamefont {R{\"{o}}sch}}, \bibinfo
  {author} {\bibfnamefont {F.}~\bibnamefont {Wagner}}, \bibinfo {author}
  {\bibfnamefont {B.}~\bibnamefont {Zielbauer}}, \ and\ \bibinfo {author}
  {\bibfnamefont {J.}~\bibnamefont {Schreiber}},\ }\href {\doibase
  10.1038/s41467-017-02663-1} {\bibfield  {journal} {\bibinfo  {journal}
  {Nature Communications}\ }\textbf {\bibinfo {volume} {9}},\ \bibinfo {pages}
  {423} (\bibinfo {year} {2018})}\BibitemShut {NoStop}%
\bibitem [{\citenamefont {Tatomirescu}\ \emph {et~al.}(2019)\citenamefont
  {Tatomirescu}, \citenamefont {Vizman},\ and\ \citenamefont
  {D'Humi{\`{e}}res}}]{Tatomirescu2019}%
  \BibitemOpen
  \bibfield  {author} {\bibinfo {author} {\bibfnamefont {D.}~\bibnamefont
  {Tatomirescu}}, \bibinfo {author} {\bibfnamefont {D.}~\bibnamefont {Vizman}},
  \ and\ \bibinfo {author} {\bibfnamefont {E.}~\bibnamefont
  {D'Humi{\`{e}}res}},\ }\href {\doibase 10.1088/1361-6587/ab469a} {\bibfield
  {journal} {\bibinfo  {journal} {Plasma Physics and Controlled Fusion}\
  }\textbf {\bibinfo {volume} {61}},\ \bibinfo {pages} {114004} (\bibinfo
  {year} {2019})}\BibitemShut {NoStop}%
\bibitem [{\citenamefont {Snavely}\ \emph {et~al.}(2000)\citenamefont
  {Snavely}, \citenamefont {Key}, \citenamefont {Hatchett}, \citenamefont
  {Cowan}, \citenamefont {Roth}, \citenamefont {Phillips}, \citenamefont
  {Stoyer}, \citenamefont {Henry}, \citenamefont {Sangster}, \citenamefont
  {Singh}, \citenamefont {Wilks}, \citenamefont {MacKinnon}, \citenamefont
  {Offenberger}, \citenamefont {Pennington}, \citenamefont {Yasuike},
  \citenamefont {Langdon}, \citenamefont {Lasinski}, \citenamefont {Johnson},
  \citenamefont {Perry},\ and\ \citenamefont
  {Campbell}}]{Snavely-IntenseProtonBeams}%
  \BibitemOpen
  \bibfield  {author} {\bibinfo {author} {\bibfnamefont {R.~A.}\ \bibnamefont
  {Snavely}}, \bibinfo {author} {\bibfnamefont {M.~H.}\ \bibnamefont {Key}},
  \bibinfo {author} {\bibfnamefont {S.~P.}\ \bibnamefont {Hatchett}}, \bibinfo
  {author} {\bibfnamefont {I.~E.}\ \bibnamefont {Cowan}}, \bibinfo {author}
  {\bibfnamefont {M.}~\bibnamefont {Roth}}, \bibinfo {author} {\bibfnamefont
  {T.~W.}\ \bibnamefont {Phillips}}, \bibinfo {author} {\bibfnamefont {M.~A.}\
  \bibnamefont {Stoyer}}, \bibinfo {author} {\bibfnamefont {E.~A.}\
  \bibnamefont {Henry}}, \bibinfo {author} {\bibfnamefont {T.~C.}\ \bibnamefont
  {Sangster}}, \bibinfo {author} {\bibfnamefont {M.~S.}\ \bibnamefont {Singh}},
  \bibinfo {author} {\bibfnamefont {S.~C.}\ \bibnamefont {Wilks}}, \bibinfo
  {author} {\bibfnamefont {A.}~\bibnamefont {MacKinnon}}, \bibinfo {author}
  {\bibfnamefont {A.}~\bibnamefont {Offenberger}}, \bibinfo {author}
  {\bibfnamefont {D.~M.}\ \bibnamefont {Pennington}}, \bibinfo {author}
  {\bibfnamefont {K.}~\bibnamefont {Yasuike}}, \bibinfo {author} {\bibfnamefont
  {A.~B.}\ \bibnamefont {Langdon}}, \bibinfo {author} {\bibfnamefont {B.~F.}\
  \bibnamefont {Lasinski}}, \bibinfo {author} {\bibfnamefont {J.}~\bibnamefont
  {Johnson}}, \bibinfo {author} {\bibfnamefont {M.~D.}\ \bibnamefont {Perry}},
  \ and\ \bibinfo {author} {\bibfnamefont {E.~M.}\ \bibnamefont {Campbell}},\
  }\href {\doibase 10.1103/PhysRevLett.85.2945} {\bibfield  {journal} {\bibinfo
   {journal} {Physical Review Letters}\ }\textbf {\bibinfo {volume} {85}},\
  \bibinfo {pages} {2945} (\bibinfo {year} {2000})}\BibitemShut {NoStop}%
\bibitem [{\citenamefont {Higginson}\ \emph {et~al.}(2018)\citenamefont
  {Higginson}, \citenamefont {Gray}, \citenamefont {King}, \citenamefont
  {Dance}, \citenamefont {Williamson}, \citenamefont {Butler}, \citenamefont
  {Wilson}, \citenamefont {Capdessus}, \citenamefont {Armstrong}, \citenamefont
  {Green}, \citenamefont {Hawkes}, \citenamefont {Martin}, \citenamefont {Wei},
  \citenamefont {Mirfayzi}, \citenamefont {Yuan}, \citenamefont {Kar},
  \citenamefont {Borghesi}, \citenamefont {Clarke}, \citenamefont {Neely},\
  and\ \citenamefont {McKenna}}]{Higginson2018}%
  \BibitemOpen
  \bibfield  {author} {\bibinfo {author} {\bibfnamefont {A.}~\bibnamefont
  {Higginson}}, \bibinfo {author} {\bibfnamefont {R.~J.}\ \bibnamefont {Gray}},
  \bibinfo {author} {\bibfnamefont {M.}~\bibnamefont {King}}, \bibinfo {author}
  {\bibfnamefont {R.~J.}\ \bibnamefont {Dance}}, \bibinfo {author}
  {\bibfnamefont {S.~D.}\ \bibnamefont {Williamson}}, \bibinfo {author}
  {\bibfnamefont {N.~M.}\ \bibnamefont {Butler}}, \bibinfo {author}
  {\bibfnamefont {R.}~\bibnamefont {Wilson}}, \bibinfo {author} {\bibfnamefont
  {R.}~\bibnamefont {Capdessus}}, \bibinfo {author} {\bibfnamefont
  {C.}~\bibnamefont {Armstrong}}, \bibinfo {author} {\bibfnamefont {J.~S.}\
  \bibnamefont {Green}}, \bibinfo {author} {\bibfnamefont {S.~J.}\ \bibnamefont
  {Hawkes}}, \bibinfo {author} {\bibfnamefont {P.}~\bibnamefont {Martin}},
  \bibinfo {author} {\bibfnamefont {W.~Q.}\ \bibnamefont {Wei}}, \bibinfo
  {author} {\bibfnamefont {S.~R.}\ \bibnamefont {Mirfayzi}}, \bibinfo {author}
  {\bibfnamefont {X.~H.}\ \bibnamefont {Yuan}}, \bibinfo {author}
  {\bibfnamefont {S.}~\bibnamefont {Kar}}, \bibinfo {author} {\bibfnamefont
  {M.}~\bibnamefont {Borghesi}}, \bibinfo {author} {\bibfnamefont {R.~J.}\
  \bibnamefont {Clarke}}, \bibinfo {author} {\bibfnamefont {D.}~\bibnamefont
  {Neely}}, \ and\ \bibinfo {author} {\bibfnamefont {P.}~\bibnamefont
  {McKenna}},\ }\href {\doibase 10.1038/s41467-018-03063-9} {\bibfield
  {journal} {\bibinfo  {journal} {Nature Communications}\ }\textbf {\bibinfo
  {volume} {9}},\ \bibinfo {pages} {724} (\bibinfo {year} {2018})}\BibitemShut
  {NoStop}%
\bibitem [{\citenamefont {Hornung}\ \emph {et~al.}(2020)\citenamefont
  {Hornung}, \citenamefont {Zobus}, \citenamefont {Boller}, \citenamefont
  {Brabetz}, \citenamefont {Eisenbarth}, \citenamefont {Kuhl}, \citenamefont
  {Major}, \citenamefont {Ohland}, \citenamefont {Zepf}, \citenamefont
  {Zielbauer},\ and\ \citenamefont {Bagnoud1}}]{Hornung2020}%
  \BibitemOpen
  \bibfield  {author} {\bibinfo {author} {\bibfnamefont {J.}~\bibnamefont
  {Hornung}}, \bibinfo {author} {\bibfnamefont {Y.}~\bibnamefont {Zobus}},
  \bibinfo {author} {\bibfnamefont {P.}~\bibnamefont {Boller}}, \bibinfo
  {author} {\bibfnamefont {C.}~\bibnamefont {Brabetz}}, \bibinfo {author}
  {\bibfnamefont {U.}~\bibnamefont {Eisenbarth}}, \bibinfo {author}
  {\bibfnamefont {T.}~\bibnamefont {Kuhl}}, \bibinfo {author} {\bibfnamefont
  {Z.}~\bibnamefont {Major}}, \bibinfo {author} {\bibfnamefont {J.~B.}\
  \bibnamefont {Ohland}}, \bibinfo {author} {\bibfnamefont {M.}~\bibnamefont
  {Zepf}}, \bibinfo {author} {\bibfnamefont {B.}~\bibnamefont {Zielbauer}}, \
  and\ \bibinfo {author} {\bibfnamefont {V.}~\bibnamefont {Bagnoud1}},\ }\href
  {\doibase 10.1017/hpl.2020.23} {\bibfield  {journal} {\bibinfo  {journal}
  {High Power Laser Science and Engineering}\ }\textbf {\bibinfo {volume}
  {8}},\ \bibinfo {pages} {1} (\bibinfo {year} {2020})}\BibitemShut {NoStop}%
\bibitem [{\citenamefont {Kraft}\ \emph {et~al.}(2010)\citenamefont {Kraft},
  \citenamefont {Richter}, \citenamefont {Zeil}, \citenamefont {Baumann},
  \citenamefont {Beyreuther}, \citenamefont {Bock}, \citenamefont {Bussmann},
  \citenamefont {Cowan}, \citenamefont {Dammene}, \citenamefont {Enghardt},
  \citenamefont {Heibig}, \citenamefont {Karsch}, \citenamefont {Kluge},
  \citenamefont {Laschinsky}, \citenamefont {Lessmann}, \citenamefont
  {Metzkes}, \citenamefont {Naumburger}, \citenamefont {Sauerbrey},
  \citenamefont {Sch{\"{u}}rer}, \citenamefont {Sobiella}, \citenamefont
  {Woithe}, \citenamefont {Schramm},\ and\ \citenamefont
  {Pawelke}}]{Kraft2010}%
  \BibitemOpen
  \bibfield  {author} {\bibinfo {author} {\bibfnamefont {S.~D.}\ \bibnamefont
  {Kraft}}, \bibinfo {author} {\bibfnamefont {C.}~\bibnamefont {Richter}},
  \bibinfo {author} {\bibfnamefont {K.}~\bibnamefont {Zeil}}, \bibinfo {author}
  {\bibfnamefont {M.}~\bibnamefont {Baumann}}, \bibinfo {author} {\bibfnamefont
  {E.}~\bibnamefont {Beyreuther}}, \bibinfo {author} {\bibfnamefont
  {S.}~\bibnamefont {Bock}}, \bibinfo {author} {\bibfnamefont {M.}~\bibnamefont
  {Bussmann}}, \bibinfo {author} {\bibfnamefont {T.~E.}\ \bibnamefont {Cowan}},
  \bibinfo {author} {\bibfnamefont {Y.}~\bibnamefont {Dammene}}, \bibinfo
  {author} {\bibfnamefont {W.}~\bibnamefont {Enghardt}}, \bibinfo {author}
  {\bibfnamefont {U.}~\bibnamefont {Heibig}}, \bibinfo {author} {\bibfnamefont
  {L.}~\bibnamefont {Karsch}}, \bibinfo {author} {\bibfnamefont
  {T.}~\bibnamefont {Kluge}}, \bibinfo {author} {\bibfnamefont
  {L.}~\bibnamefont {Laschinsky}}, \bibinfo {author} {\bibfnamefont
  {E.}~\bibnamefont {Lessmann}}, \bibinfo {author} {\bibfnamefont
  {J.}~\bibnamefont {Metzkes}}, \bibinfo {author} {\bibfnamefont
  {D.}~\bibnamefont {Naumburger}}, \bibinfo {author} {\bibfnamefont
  {R.}~\bibnamefont {Sauerbrey}}, \bibinfo {author} {\bibfnamefont
  {M.}~\bibnamefont {Sch{\"{u}}rer}}, \bibinfo {author} {\bibfnamefont
  {M.}~\bibnamefont {Sobiella}}, \bibinfo {author} {\bibfnamefont
  {J.}~\bibnamefont {Woithe}}, \bibinfo {author} {\bibfnamefont
  {U.}~\bibnamefont {Schramm}}, \ and\ \bibinfo {author} {\bibfnamefont
  {J.}~\bibnamefont {Pawelke}},\ }\href {\doibase
  10.1088/1367-2630/12/8/085003} {\bibfield  {journal} {\bibinfo  {journal}
  {New Journal of Physics}\ }\textbf {\bibinfo {volume} {12}},\ \bibinfo
  {pages} {085003} (\bibinfo {year} {2010})}\BibitemShut {NoStop}%
\bibitem [{\citenamefont {Karsch}\ \emph {et~al.}(2017)\citenamefont {Karsch},
  \citenamefont {Beyreuther}, \citenamefont {Enghardt}, \citenamefont {Gotz},
  \citenamefont {Masood}, \citenamefont {Schramm}, \citenamefont {Zeil},\ and\
  \citenamefont {Pawelke}}]{Karsch2017}%
  \BibitemOpen
  \bibfield  {author} {\bibinfo {author} {\bibfnamefont {L.}~\bibnamefont
  {Karsch}}, \bibinfo {author} {\bibfnamefont {E.}~\bibnamefont {Beyreuther}},
  \bibinfo {author} {\bibfnamefont {W.}~\bibnamefont {Enghardt}}, \bibinfo
  {author} {\bibfnamefont {M.}~\bibnamefont {Gotz}}, \bibinfo {author}
  {\bibfnamefont {U.}~\bibnamefont {Masood}}, \bibinfo {author} {\bibfnamefont
  {U.}~\bibnamefont {Schramm}}, \bibinfo {author} {\bibfnamefont
  {K.}~\bibnamefont {Zeil}}, \ and\ \bibinfo {author} {\bibfnamefont
  {J.}~\bibnamefont {Pawelke}},\ }\href {\doibase
  10.1080/0284186X.2017.1355111} {\bibfield  {journal} {\bibinfo  {journal}
  {Acta Oncologica}\ }\textbf {\bibinfo {volume} {56}},\ \bibinfo {pages}
  {1359} (\bibinfo {year} {2017})}\BibitemShut {NoStop}%
\bibitem [{\citenamefont {Maksimchuk}\ \emph {et~al.}(2000)\citenamefont
  {Maksimchuk}, \citenamefont {Gu}, \citenamefont {Flippo}, \citenamefont
  {Umstadter},\ and\ \citenamefont {Bychenkov}}]{Maksimchuck2000}%
  \BibitemOpen
  \bibfield  {author} {\bibinfo {author} {\bibfnamefont {A.}~\bibnamefont
  {Maksimchuk}}, \bibinfo {author} {\bibfnamefont {S.}~\bibnamefont {Gu}},
  \bibinfo {author} {\bibfnamefont {K.}~\bibnamefont {Flippo}}, \bibinfo
  {author} {\bibfnamefont {D.}~\bibnamefont {Umstadter}}, \ and\ \bibinfo
  {author} {\bibfnamefont {A.~Y.}\ \bibnamefont {Bychenkov}},\ }\href {\doibase
  10.1103/PhysRevLett.84.4108} {\bibfield  {journal} {\bibinfo  {journal}
  {Physical Review Letters}\ }\textbf {\bibinfo {volume} {84}},\ \bibinfo
  {pages} {4108} (\bibinfo {year} {2000})}\BibitemShut {NoStop}%
\bibitem [{\citenamefont {Hatchett}\ \emph {et~al.}(2000)\citenamefont
  {Hatchett}, \citenamefont {Brown}, \citenamefont {Cowan}, \citenamefont
  {Henry}, \citenamefont {Johnson}, \citenamefont {Key}, \citenamefont {Koch},
  \citenamefont {Langdon}, \citenamefont {Lasinski}, \citenamefont {Lee},
  \citenamefont {Mackinnon}, \citenamefont {Pennington}, \citenamefont {Perry},
  \citenamefont {Phillips}, \citenamefont {Roth}, \citenamefont {Sangster},
  \citenamefont {Singh}, \citenamefont {Snavely}, \citenamefont {Stoyer},
  \citenamefont {Wilks},\ and\ \citenamefont
  {Yasuike}}]{Hatchett-ElectronPhotonAndIons}%
  \BibitemOpen
  \bibfield  {author} {\bibinfo {author} {\bibfnamefont {S.~P.}\ \bibnamefont
  {Hatchett}}, \bibinfo {author} {\bibfnamefont {C.~G.}\ \bibnamefont {Brown}},
  \bibinfo {author} {\bibfnamefont {T.~E.}\ \bibnamefont {Cowan}}, \bibinfo
  {author} {\bibfnamefont {E.~A.}\ \bibnamefont {Henry}}, \bibinfo {author}
  {\bibfnamefont {J.~S.}\ \bibnamefont {Johnson}}, \bibinfo {author}
  {\bibfnamefont {M.~H.}\ \bibnamefont {Key}}, \bibinfo {author} {\bibfnamefont
  {J.~A.}\ \bibnamefont {Koch}}, \bibinfo {author} {\bibfnamefont {A.~B.}\
  \bibnamefont {Langdon}}, \bibinfo {author} {\bibfnamefont {B.~F.}\
  \bibnamefont {Lasinski}}, \bibinfo {author} {\bibfnamefont {R.~W.}\
  \bibnamefont {Lee}}, \bibinfo {author} {\bibfnamefont {A.~J.}\ \bibnamefont
  {Mackinnon}}, \bibinfo {author} {\bibfnamefont {D.~M.}\ \bibnamefont
  {Pennington}}, \bibinfo {author} {\bibfnamefont {M.~D.}\ \bibnamefont
  {Perry}}, \bibinfo {author} {\bibfnamefont {T.~W.}\ \bibnamefont {Phillips}},
  \bibinfo {author} {\bibfnamefont {M.}~\bibnamefont {Roth}}, \bibinfo {author}
  {\bibfnamefont {T.~C.}\ \bibnamefont {Sangster}}, \bibinfo {author}
  {\bibfnamefont {M.~S.}\ \bibnamefont {Singh}}, \bibinfo {author}
  {\bibfnamefont {R.~A.}\ \bibnamefont {Snavely}}, \bibinfo {author}
  {\bibfnamefont {M.~A.}\ \bibnamefont {Stoyer}}, \bibinfo {author}
  {\bibfnamefont {S.~C.}\ \bibnamefont {Wilks}}, \ and\ \bibinfo {author}
  {\bibfnamefont {K.}~\bibnamefont {Yasuike}},\ }\href {\doibase
  10.1063/1.874030} {\bibfield  {journal} {\bibinfo  {journal} {Physics of
  Plasmas}\ }\textbf {\bibinfo {volume} {7}},\ \bibinfo {pages} {2076}
  (\bibinfo {year} {2000})}\BibitemShut {NoStop}%
\bibitem [{\citenamefont {Wilks}\ \emph {et~al.}(2001)\citenamefont {Wilks},
  \citenamefont {Langdon}, \citenamefont {Cowan}, \citenamefont {Roth},
  \citenamefont {Singh}, \citenamefont {Hatchett}, \citenamefont {Key},
  \citenamefont {Pennington}, \citenamefont {MacKinnon},\ and\ \citenamefont
  {Snavely}}]{wilks}%
  \BibitemOpen
  \bibfield  {author} {\bibinfo {author} {\bibfnamefont {S.~C.}\ \bibnamefont
  {Wilks}}, \bibinfo {author} {\bibfnamefont {A.~B.}\ \bibnamefont {Langdon}},
  \bibinfo {author} {\bibfnamefont {T.~E.}\ \bibnamefont {Cowan}}, \bibinfo
  {author} {\bibfnamefont {M.}~\bibnamefont {Roth}}, \bibinfo {author}
  {\bibfnamefont {M.}~\bibnamefont {Singh}}, \bibinfo {author} {\bibfnamefont
  {S.}~\bibnamefont {Hatchett}}, \bibinfo {author} {\bibfnamefont {M.~H.}\
  \bibnamefont {Key}}, \bibinfo {author} {\bibfnamefont {D.}~\bibnamefont
  {Pennington}}, \bibinfo {author} {\bibfnamefont {A.}~\bibnamefont
  {MacKinnon}}, \ and\ \bibinfo {author} {\bibfnamefont {R.~A.}\ \bibnamefont
  {Snavely}},\ }\href {\doibase 10.1063/1.1333697} {\bibfield  {journal}
  {\bibinfo  {journal} {Physics of Plasmas}\ }\textbf {\bibinfo {volume} {8}},\
  \bibinfo {pages} {542} (\bibinfo {year} {2001})}\BibitemShut {NoStop}%
\bibitem [{\citenamefont {D'Hum{\`{i}}res}\ \emph {et~al.}(2005)\citenamefont
  {D'Hum{\`{i}}res}, \citenamefont {Lefebvre}, \citenamefont {Gremillet},\ and\
  \citenamefont {Malka}}]{DHumieres2005}%
  \BibitemOpen
  \bibfield  {author} {\bibinfo {author} {\bibfnamefont {E.}~\bibnamefont
  {D'Hum{\`{i}}res}}, \bibinfo {author} {\bibfnamefont {E.}~\bibnamefont
  {Lefebvre}}, \bibinfo {author} {\bibfnamefont {L.}~\bibnamefont {Gremillet}},
  \ and\ \bibinfo {author} {\bibfnamefont {V.}~\bibnamefont {Malka}},\ }\href
  {\doibase 10.1063/1.1927097} {\bibfield  {journal} {\bibinfo  {journal}
  {Physics of Plasmas}\ }\textbf {\bibinfo {volume} {12}},\ \bibinfo {pages}
  {1} (\bibinfo {year} {2005})}\BibitemShut {NoStop}%
\bibitem [{\citenamefont {Yan}\ \emph {et~al.}(2010)\citenamefont {Yan},
  \citenamefont {Tajima}, \citenamefont {Hegelich}, \citenamefont {Yin},\ and\
  \citenamefont {Habs}}]{Yan2010}%
  \BibitemOpen
  \bibfield  {author} {\bibinfo {author} {\bibfnamefont {X.~Q.}\ \bibnamefont
  {Yan}}, \bibinfo {author} {\bibfnamefont {T.}~\bibnamefont {Tajima}},
  \bibinfo {author} {\bibfnamefont {M.}~\bibnamefont {Hegelich}}, \bibinfo
  {author} {\bibfnamefont {L.}~\bibnamefont {Yin}}, \ and\ \bibinfo {author}
  {\bibfnamefont {D.}~\bibnamefont {Habs}},\ }\href {\doibase
  10.1007/s00340-009-3707-5} {\bibfield  {journal} {\bibinfo  {journal}
  {Applied Physics B: Lasers and Optics}\ }\textbf {\bibinfo {volume} {98}},\
  \bibinfo {pages} {711} (\bibinfo {year} {2010})}\BibitemShut {NoStop}%
\bibitem [{\citenamefont {Mishra}\ \emph {et~al.}(2018)\citenamefont {Mishra},
  \citenamefont {Fiuza},\ and\ \citenamefont {Glenzer}}]{Mishra2018}%
  \BibitemOpen
  \bibfield  {author} {\bibinfo {author} {\bibfnamefont {R.}~\bibnamefont
  {Mishra}}, \bibinfo {author} {\bibfnamefont {F.}~\bibnamefont {Fiuza}}, \
  and\ \bibinfo {author} {\bibfnamefont {S.}~\bibnamefont {Glenzer}},\ }\href
  {\doibase 10.1088/1367-2630/aab8db} {\bibfield  {journal} {\bibinfo
  {journal} {New Journal of Physics}\ }\textbf {\bibinfo {volume} {20}},\
  \bibinfo {pages} {043047} (\bibinfo {year} {2018})}\BibitemShut {NoStop}%
\bibitem [{\citenamefont {Poole}\ \emph {et~al.}(2018)\citenamefont {Poole},
  \citenamefont {Obst}, \citenamefont {Cochran}, \citenamefont {Metzkes},
  \citenamefont {Schlenvoigt}, \citenamefont {Prencipe}, \citenamefont {Kluge},
  \citenamefont {Cowan}, \citenamefont {Schramm}, \citenamefont {Schumacher},\
  and\ \citenamefont {Zeil}}]{Poole2018}%
  \BibitemOpen
  \bibfield  {author} {\bibinfo {author} {\bibfnamefont {P.~L.}\ \bibnamefont
  {Poole}}, \bibinfo {author} {\bibfnamefont {L.}~\bibnamefont {Obst}},
  \bibinfo {author} {\bibfnamefont {G.~E.}\ \bibnamefont {Cochran}}, \bibinfo
  {author} {\bibfnamefont {J.}~\bibnamefont {Metzkes}}, \bibinfo {author}
  {\bibfnamefont {H.~P.}\ \bibnamefont {Schlenvoigt}}, \bibinfo {author}
  {\bibfnamefont {I.}~\bibnamefont {Prencipe}}, \bibinfo {author}
  {\bibfnamefont {T.}~\bibnamefont {Kluge}}, \bibinfo {author} {\bibfnamefont
  {T.}~\bibnamefont {Cowan}}, \bibinfo {author} {\bibfnamefont
  {U.}~\bibnamefont {Schramm}}, \bibinfo {author} {\bibfnamefont {D.~W.}\
  \bibnamefont {Schumacher}}, \ and\ \bibinfo {author} {\bibfnamefont
  {K.}~\bibnamefont {Zeil}},\ }\href {\doibase 10.1088/1367-2630/aa9d47}
  {\bibfield  {journal} {\bibinfo  {journal} {New Journal of Physics}\ }\textbf
  {\bibinfo {volume} {20}},\ \bibinfo {pages} {013019} (\bibinfo {year}
  {2018})}\BibitemShut {NoStop}%
\bibitem [{\citenamefont {Esirkepov}\ \emph {et~al.}(2004)\citenamefont
  {Esirkepov}, \citenamefont {Borghesi}, \citenamefont {Bulanov}, \citenamefont
  {Mourou},\ and\ \citenamefont {Tajima}}]{Esirkepov-LaserPiston}%
  \BibitemOpen
  \bibfield  {author} {\bibinfo {author} {\bibfnamefont {T.}~\bibnamefont
  {Esirkepov}}, \bibinfo {author} {\bibfnamefont {M.}~\bibnamefont {Borghesi}},
  \bibinfo {author} {\bibfnamefont {S.~V.}\ \bibnamefont {Bulanov}}, \bibinfo
  {author} {\bibfnamefont {G.}~\bibnamefont {Mourou}}, \ and\ \bibinfo {author}
  {\bibfnamefont {T.}~\bibnamefont {Tajima}},\ }\href {\doibase
  10.1103/PhysRevLett.92.175003} {\bibfield  {journal} {\bibinfo  {journal}
  {Physical Review Letters}\ }\textbf {\bibinfo {volume} {92}},\ \bibinfo
  {pages} {175003} (\bibinfo {year} {2004})}\BibitemShut {NoStop}%
\bibitem [{\citenamefont {Robinson}\ \emph {et~al.}(2009)\citenamefont
  {Robinson}, \citenamefont {Gibbon}, \citenamefont {Zepf}, \citenamefont
  {Kar}, \citenamefont {Evans},\ and\ \citenamefont {Bellei}}]{Robinson2009}%
  \BibitemOpen
  \bibfield  {author} {\bibinfo {author} {\bibfnamefont {A.~P.}\ \bibnamefont
  {Robinson}}, \bibinfo {author} {\bibfnamefont {P.}~\bibnamefont {Gibbon}},
  \bibinfo {author} {\bibfnamefont {M.}~\bibnamefont {Zepf}}, \bibinfo {author}
  {\bibfnamefont {S.}~\bibnamefont {Kar}}, \bibinfo {author} {\bibfnamefont
  {R.~G.}\ \bibnamefont {Evans}}, \ and\ \bibinfo {author} {\bibfnamefont
  {C.}~\bibnamefont {Bellei}},\ }\href {\doibase 10.1088/0741-3335/51/2/024004}
  {\bibfield  {journal} {\bibinfo  {journal} {Plasma Physics and Controlled
  Fusion}\ }\textbf {\bibinfo {volume} {51}},\ \bibinfo {pages} {024004}
  (\bibinfo {year} {2009})}\BibitemShut {NoStop}%
\bibitem [{\citenamefont {Simmons}\ and\ \citenamefont
  {McInnes}(1993)}]{Simmons1993}%
  \BibitemOpen
  \bibfield  {author} {\bibinfo {author} {\bibfnamefont {J.~F.~L.}\
  \bibnamefont {Simmons}}\ and\ \bibinfo {author} {\bibfnamefont {C.~R.}\
  \bibnamefont {McInnes}},\ }\href {\doibase 10.1119/1.17291} {\bibfield
  {journal} {\bibinfo  {journal} {American Journal of Physics}\ }\textbf
  {\bibinfo {volume} {61}},\ \bibinfo {pages} {205} (\bibinfo {year}
  {1993})}\BibitemShut {NoStop}%
\bibitem [{\citenamefont {MacChi}\ \emph {et~al.}(2009)\citenamefont {MacChi},
  \citenamefont {Veghini},\ and\ \citenamefont {Pegoraro}}]{macchi2009light}%
  \BibitemOpen
  \bibfield  {author} {\bibinfo {author} {\bibfnamefont {A.}~\bibnamefont
  {MacChi}}, \bibinfo {author} {\bibfnamefont {S.}~\bibnamefont {Veghini}}, \
  and\ \bibinfo {author} {\bibfnamefont {F.}~\bibnamefont {Pegoraro}},\ }\href
  {\doibase 10.1103/PhysRevLett.103.085003} {\bibfield  {journal} {\bibinfo
  {journal} {Physical Review Letters}\ }\textbf {\bibinfo {volume} {103}},\
  \bibinfo {pages} {085003} (\bibinfo {year} {2009})}\BibitemShut {NoStop}%
\bibitem [{\citenamefont {Pegoraro}\ and\ \citenamefont
  {Bulanov}(2007)}]{Pegoraro2007}%
  \BibitemOpen
  \bibfield  {author} {\bibinfo {author} {\bibfnamefont {F.}~\bibnamefont
  {Pegoraro}}\ and\ \bibinfo {author} {\bibfnamefont {S.~V.}\ \bibnamefont
  {Bulanov}},\ }\href {\doibase 10.1103/PhysRevLett.99.065002} {\bibfield
  {journal} {\bibinfo  {journal} {Physical Review Letters}\ }\textbf {\bibinfo
  {volume} {99}},\ \bibinfo {pages} {065002} (\bibinfo {year}
  {2007})}\BibitemShut {NoStop}%
\bibitem [{\citenamefont {Palmer}\ \emph {et~al.}(2012)\citenamefont {Palmer},
  \citenamefont {Schreiber}, \citenamefont {Nagel}, \citenamefont {Dover},
  \citenamefont {Bellei}, \citenamefont {Beg}, \citenamefont {Bott},
  \citenamefont {Clarke}, \citenamefont {Dangor}, \citenamefont {Hassan},
  \citenamefont {Hilz}, \citenamefont {Jung}, \citenamefont {Kneip},
  \citenamefont {Mangles}, \citenamefont {Lancaster}, \citenamefont {Rehman},
  \citenamefont {Robinson}, \citenamefont {Spindloe}, \citenamefont {Szerypo},
  \citenamefont {Tatarakis}, \citenamefont {Yeung}, \citenamefont {Zepf},\ and\
  \citenamefont {Najmudin}}]{Palmer2012}%
  \BibitemOpen
  \bibfield  {author} {\bibinfo {author} {\bibfnamefont {C.~A.}\ \bibnamefont
  {Palmer}}, \bibinfo {author} {\bibfnamefont {J.}~\bibnamefont {Schreiber}},
  \bibinfo {author} {\bibfnamefont {S.~R.}\ \bibnamefont {Nagel}}, \bibinfo
  {author} {\bibfnamefont {N.~P.}\ \bibnamefont {Dover}}, \bibinfo {author}
  {\bibfnamefont {C.}~\bibnamefont {Bellei}}, \bibinfo {author} {\bibfnamefont
  {F.~N.}\ \bibnamefont {Beg}}, \bibinfo {author} {\bibfnamefont
  {S.}~\bibnamefont {Bott}}, \bibinfo {author} {\bibfnamefont {R.~J.}\
  \bibnamefont {Clarke}}, \bibinfo {author} {\bibfnamefont {A.~E.}\
  \bibnamefont {Dangor}}, \bibinfo {author} {\bibfnamefont {S.~M.}\
  \bibnamefont {Hassan}}, \bibinfo {author} {\bibfnamefont {P.}~\bibnamefont
  {Hilz}}, \bibinfo {author} {\bibfnamefont {D.}~\bibnamefont {Jung}}, \bibinfo
  {author} {\bibfnamefont {S.}~\bibnamefont {Kneip}}, \bibinfo {author}
  {\bibfnamefont {S.~P.}\ \bibnamefont {Mangles}}, \bibinfo {author}
  {\bibfnamefont {K.~L.}\ \bibnamefont {Lancaster}}, \bibinfo {author}
  {\bibfnamefont {A.}~\bibnamefont {Rehman}}, \bibinfo {author} {\bibfnamefont
  {A.~P.}\ \bibnamefont {Robinson}}, \bibinfo {author} {\bibfnamefont
  {C.}~\bibnamefont {Spindloe}}, \bibinfo {author} {\bibfnamefont
  {J.}~\bibnamefont {Szerypo}}, \bibinfo {author} {\bibfnamefont
  {M.}~\bibnamefont {Tatarakis}}, \bibinfo {author} {\bibfnamefont
  {M.}~\bibnamefont {Yeung}}, \bibinfo {author} {\bibfnamefont
  {M.}~\bibnamefont {Zepf}}, \ and\ \bibinfo {author} {\bibfnamefont
  {Z.}~\bibnamefont {Najmudin}},\ }\href {\doibase
  10.1103/PhysRevLett.108.225002} {\bibfield  {journal} {\bibinfo  {journal}
  {Physical Review Letters}\ }\textbf {\bibinfo {volume} {108}},\ \bibinfo
  {pages} {225002} (\bibinfo {year} {2012})}\BibitemShut {NoStop}%
\bibitem [{\citenamefont {Obst}\ \emph {et~al.}(2017)\citenamefont {Obst},
  \citenamefont {G{\"{o}}de}, \citenamefont {Rehwald}, \citenamefont {Brack},
  \citenamefont {Branco}, \citenamefont {Bock}, \citenamefont {Bussmann},
  \citenamefont {Cowan}, \citenamefont {Curry}, \citenamefont {Fiuza},
  \citenamefont {Gauthier}, \citenamefont {Gebhardt}, \citenamefont {Helbig},
  \citenamefont {Huebl}, \citenamefont {H{\"{u}}bner}, \citenamefont {Irman},
  \citenamefont {Kazak}, \citenamefont {Kim}, \citenamefont {Kluge},
  \citenamefont {Kraft}, \citenamefont {Loeser}, \citenamefont {Metzkes},
  \citenamefont {Mishra}, \citenamefont {R{\"{o}}del}, \citenamefont
  {Schlenvoigt}, \citenamefont {Siebold}, \citenamefont {Tiggesb{\"{a}}umker},
  \citenamefont {Wolter}, \citenamefont {Ziegler}, \citenamefont {Schramm},
  \citenamefont {Glenzer},\ and\ \citenamefont {Zeil}}]{Obst2017}%
  \BibitemOpen
  \bibfield  {author} {\bibinfo {author} {\bibfnamefont {L.}~\bibnamefont
  {Obst}}, \bibinfo {author} {\bibfnamefont {S.}~\bibnamefont {G{\"{o}}de}},
  \bibinfo {author} {\bibfnamefont {M.}~\bibnamefont {Rehwald}}, \bibinfo
  {author} {\bibfnamefont {F.~E.}\ \bibnamefont {Brack}}, \bibinfo {author}
  {\bibfnamefont {J.}~\bibnamefont {Branco}}, \bibinfo {author} {\bibfnamefont
  {S.}~\bibnamefont {Bock}}, \bibinfo {author} {\bibfnamefont {M.}~\bibnamefont
  {Bussmann}}, \bibinfo {author} {\bibfnamefont {T.~E.}\ \bibnamefont {Cowan}},
  \bibinfo {author} {\bibfnamefont {C.~B.}\ \bibnamefont {Curry}}, \bibinfo
  {author} {\bibfnamefont {F.}~\bibnamefont {Fiuza}}, \bibinfo {author}
  {\bibfnamefont {M.}~\bibnamefont {Gauthier}}, \bibinfo {author}
  {\bibfnamefont {R.}~\bibnamefont {Gebhardt}}, \bibinfo {author}
  {\bibfnamefont {U.}~\bibnamefont {Helbig}}, \bibinfo {author} {\bibfnamefont
  {A.}~\bibnamefont {Huebl}}, \bibinfo {author} {\bibfnamefont
  {U.}~\bibnamefont {H{\"{u}}bner}}, \bibinfo {author} {\bibfnamefont
  {A.}~\bibnamefont {Irman}}, \bibinfo {author} {\bibfnamefont
  {L.}~\bibnamefont {Kazak}}, \bibinfo {author} {\bibfnamefont {J.~B.}\
  \bibnamefont {Kim}}, \bibinfo {author} {\bibfnamefont {T.}~\bibnamefont
  {Kluge}}, \bibinfo {author} {\bibfnamefont {S.}~\bibnamefont {Kraft}},
  \bibinfo {author} {\bibfnamefont {M.}~\bibnamefont {Loeser}}, \bibinfo
  {author} {\bibfnamefont {J.}~\bibnamefont {Metzkes}}, \bibinfo {author}
  {\bibfnamefont {R.}~\bibnamefont {Mishra}}, \bibinfo {author} {\bibfnamefont
  {C.}~\bibnamefont {R{\"{o}}del}}, \bibinfo {author} {\bibfnamefont {H.~P.}\
  \bibnamefont {Schlenvoigt}}, \bibinfo {author} {\bibfnamefont
  {M.}~\bibnamefont {Siebold}}, \bibinfo {author} {\bibfnamefont
  {J.}~\bibnamefont {Tiggesb{\"{a}}umker}}, \bibinfo {author} {\bibfnamefont
  {S.}~\bibnamefont {Wolter}}, \bibinfo {author} {\bibfnamefont
  {T.}~\bibnamefont {Ziegler}}, \bibinfo {author} {\bibfnamefont
  {U.}~\bibnamefont {Schramm}}, \bibinfo {author} {\bibfnamefont {S.~H.}\
  \bibnamefont {Glenzer}}, \ and\ \bibinfo {author} {\bibfnamefont
  {K.}~\bibnamefont {Zeil}},\ }\href {\doibase 10.1038/s41598-017-10589-3}
  {\bibfield  {journal} {\bibinfo  {journal} {Scientific Reports}\ }\textbf
  {\bibinfo {volume} {7}},\ \bibinfo {pages} {10248} (\bibinfo {year}
  {2017})}\BibitemShut {NoStop}%
\bibitem [{\citenamefont {Curry}\ \emph {et~al.}(2020)\citenamefont {Curry},
  \citenamefont {Schoenwaelder}, \citenamefont {Goede}, \citenamefont {Kim},
  \citenamefont {Rehwald}, \citenamefont {Treffert}, \citenamefont {Zeil},
  \citenamefont {Glenzer},\ and\ \citenamefont {Gauthier}}]{Curry2020}%
  \BibitemOpen
  \bibfield  {author} {\bibinfo {author} {\bibfnamefont {C.~B.}\ \bibnamefont
  {Curry}}, \bibinfo {author} {\bibfnamefont {C.}~\bibnamefont
  {Schoenwaelder}}, \bibinfo {author} {\bibfnamefont {S.}~\bibnamefont
  {Goede}}, \bibinfo {author} {\bibfnamefont {J.~B.}\ \bibnamefont {Kim}},
  \bibinfo {author} {\bibfnamefont {M.}~\bibnamefont {Rehwald}}, \bibinfo
  {author} {\bibfnamefont {F.}~\bibnamefont {Treffert}}, \bibinfo {author}
  {\bibfnamefont {K.}~\bibnamefont {Zeil}}, \bibinfo {author} {\bibfnamefont
  {S.~H.}\ \bibnamefont {Glenzer}}, \ and\ \bibinfo {author} {\bibfnamefont
  {M.}~\bibnamefont {Gauthier}},\ }\href {\doibase 10.3791/61130} {\bibfield
  {journal} {\bibinfo  {journal} {Journal of Visualized Experiments}\ }\textbf
  {\bibinfo {volume} {2020}} (\bibinfo {year} {2020}),\
  10.3791/61130}\BibitemShut {NoStop}%
\bibitem [{\citenamefont {Brantov}\ \emph {et~al.}(2016)\citenamefont
  {Brantov}, \citenamefont {Govras}, \citenamefont {Kovalev},\ and\
  \citenamefont {Bychenkov}}]{Brantov2016a}%
  \BibitemOpen
  \bibfield  {author} {\bibinfo {author} {\bibfnamefont {A.~V.}\ \bibnamefont
  {Brantov}}, \bibinfo {author} {\bibfnamefont {E.~A.}\ \bibnamefont {Govras}},
  \bibinfo {author} {\bibfnamefont {V.~F.}\ \bibnamefont {Kovalev}}, \ and\
  \bibinfo {author} {\bibfnamefont {V.~Y.}\ \bibnamefont {Bychenkov}},\ }\href
  {\doibase 10.1103/PhysRevLett.116.085004} {\bibfield  {journal} {\bibinfo
  {journal} {Physical Review Letters}\ }\textbf {\bibinfo {volume} {116}},\
  \bibinfo {pages} {085004} (\bibinfo {year} {2016})}\BibitemShut {NoStop}%
\bibitem [{\citenamefont {Bychenkov}\ \emph {et~al.}(2016)\citenamefont
  {Bychenkov}, \citenamefont {Govras},\ and\ \citenamefont
  {Brantov}}]{Bychenkov2016}%
  \BibitemOpen
  \bibfield  {author} {\bibinfo {author} {\bibfnamefont {V.~Y.}\ \bibnamefont
  {Bychenkov}}, \bibinfo {author} {\bibfnamefont {E.~A.}\ \bibnamefont
  {Govras}}, \ and\ \bibinfo {author} {\bibfnamefont {A.~V.}\ \bibnamefont
  {Brantov}},\ }\href {\doibase 10.1134/S0021364016210086} {\bibfield
  {journal} {\bibinfo  {journal} {JETP Letters}\ }\textbf {\bibinfo {volume}
  {104}},\ \bibinfo {pages} {618} (\bibinfo {year} {2016})}\BibitemShut
  {NoStop}%
\bibitem [{\citenamefont {Brantov}\ \emph
  {et~al.}(2017{\natexlab{a}})\citenamefont {Brantov}, \citenamefont
  {Obraztsova}, \citenamefont {Chuvilin}, \citenamefont {Obraztsova},\ and\
  \citenamefont {Bychenkov}}]{Brantov2017a}%
  \BibitemOpen
  \bibfield  {author} {\bibinfo {author} {\bibfnamefont {A.~V.}\ \bibnamefont
  {Brantov}}, \bibinfo {author} {\bibfnamefont {E.~A.}\ \bibnamefont
  {Obraztsova}}, \bibinfo {author} {\bibfnamefont {A.~L.}\ \bibnamefont
  {Chuvilin}}, \bibinfo {author} {\bibfnamefont {E.~D.}\ \bibnamefont
  {Obraztsova}}, \ and\ \bibinfo {author} {\bibfnamefont {V.~Y.}\ \bibnamefont
  {Bychenkov}},\ }\href {\doibase 10.1103/PhysRevAccelBeams.20.061301}
  {\bibfield  {journal} {\bibinfo  {journal} {Physical Review Accelerators and
  Beams}\ }\textbf {\bibinfo {volume} {20}},\ \bibinfo {pages} {1} (\bibinfo
  {year} {2017}{\natexlab{a}})}\BibitemShut {NoStop}%
\bibitem [{\citenamefont {Brantov}\ \emph
  {et~al.}(2017{\natexlab{b}})\citenamefont {Brantov}, \citenamefont
  {Ksenofontov},\ and\ \citenamefont {Bychenkov}}]{Brantov2017b}%
  \BibitemOpen
  \bibfield  {author} {\bibinfo {author} {\bibfnamefont {A.~V.}\ \bibnamefont
  {Brantov}}, \bibinfo {author} {\bibfnamefont {P.~A.}\ \bibnamefont
  {Ksenofontov}}, \ and\ \bibinfo {author} {\bibfnamefont {V.~Y.}\ \bibnamefont
  {Bychenkov}},\ }\href {\doibase 10.1063/1.5003883} {\bibfield  {journal}
  {\bibinfo  {journal} {Physics of Plasmas}\ }\textbf {\bibinfo {volume} {24}}
  (\bibinfo {year} {2017}{\natexlab{b}}),\ 10.1063/1.5003883}\BibitemShut
  {NoStop}%
\bibitem [{\citenamefont {Brantov}\ and\ \citenamefont
  {Bychenkov}(2017)}]{Brantov2017c}%
  \BibitemOpen
  \bibfield  {author} {\bibinfo {author} {\bibfnamefont {A.~V.}\ \bibnamefont
  {Brantov}}\ and\ \bibinfo {author} {\bibfnamefont {V.~Y.}\ \bibnamefont
  {Bychenkov}},\ }\href {\doibase 10.1088/1361-6587/aa5862} {\bibfield
  {journal} {\bibinfo  {journal} {Plasma Physics and Controlled Fusion}\
  }\textbf {\bibinfo {volume} {59}},\ \bibinfo {pages} {1} (\bibinfo {year}
  {2017})}\BibitemShut {NoStop}%
\bibitem [{\citenamefont {Liu}\ \emph {et~al.}(2020)\citenamefont {Liu},
  \citenamefont {Meyer-Ter-Vehn}, \citenamefont {Ruhl},\ and\ \citenamefont
  {Zepf}}]{Liu2020}%
  \BibitemOpen
  \bibfield  {author} {\bibinfo {author} {\bibfnamefont {B.}~\bibnamefont
  {Liu}}, \bibinfo {author} {\bibfnamefont {J.}~\bibnamefont {Meyer-Ter-Vehn}},
  \bibinfo {author} {\bibfnamefont {H.}~\bibnamefont {Ruhl}}, \ and\ \bibinfo
  {author} {\bibfnamefont {M.}~\bibnamefont {Zepf}},\ }\href {\doibase
  10.1088/1361-6587/ab98e0} {\bibfield  {journal} {\bibinfo  {journal} {Plasma
  Physics and Controlled Fusion}\ }\textbf {\bibinfo {volume} {62}},\ \bibinfo
  {pages} {085014} (\bibinfo {year} {2020})}\BibitemShut {NoStop}%
\bibitem [{\citenamefont {Derouillat}\ \emph {et~al.}(2018)\citenamefont
  {Derouillat}, \citenamefont {Beck}, \citenamefont {P{\'{e}}rez},
  \citenamefont {Vinci}, \citenamefont {Chiaramello}, \citenamefont {Grassi},
  \citenamefont {Fl{\'{e}}}, \citenamefont {Bouchard}, \citenamefont
  {Plotnikov}, \citenamefont {Aunai}, \citenamefont {Dargent}, \citenamefont
  {Riconda},\ and\ \citenamefont {Grech}}]{Derouillat2018}%
  \BibitemOpen
  \bibfield  {author} {\bibinfo {author} {\bibfnamefont {J.}~\bibnamefont
  {Derouillat}}, \bibinfo {author} {\bibfnamefont {A.}~\bibnamefont {Beck}},
  \bibinfo {author} {\bibfnamefont {F.}~\bibnamefont {P{\'{e}}rez}}, \bibinfo
  {author} {\bibfnamefont {T.}~\bibnamefont {Vinci}}, \bibinfo {author}
  {\bibfnamefont {M.}~\bibnamefont {Chiaramello}}, \bibinfo {author}
  {\bibfnamefont {A.}~\bibnamefont {Grassi}}, \bibinfo {author} {\bibfnamefont
  {M.}~\bibnamefont {Fl{\'{e}}}}, \bibinfo {author} {\bibfnamefont
  {G.}~\bibnamefont {Bouchard}}, \bibinfo {author} {\bibfnamefont
  {I.}~\bibnamefont {Plotnikov}}, \bibinfo {author} {\bibfnamefont
  {N.}~\bibnamefont {Aunai}}, \bibinfo {author} {\bibfnamefont
  {J.}~\bibnamefont {Dargent}}, \bibinfo {author} {\bibfnamefont
  {C.}~\bibnamefont {Riconda}}, \ and\ \bibinfo {author} {\bibfnamefont
  {M.}~\bibnamefont {Grech}},\ }\href {\doibase 10.1016/j.cpc.2017.09.024}
  {\bibfield  {journal} {\bibinfo  {journal} {Computer Physics Communications}\
  }\textbf {\bibinfo {volume} {222}},\ \bibinfo {pages} {351} (\bibinfo {year}
  {2018})}\BibitemShut {NoStop}%
\bibitem [{goe(2021)}]{goethel2021scripts}%
  \BibitemOpen
  \href {\doibase 10.14278/rodare.1199} {\enquote {\bibinfo {title} {{Input
  files and model implementation, 10.14278/rodare.1199}},}\ } (\bibinfo {year}
  {2021})\BibitemShut {NoStop}%
\bibitem [{\citenamefont {Liu}\ \emph {et~al.}(2016)\citenamefont {Liu},
  \citenamefont {Meyer-Ter-Vehn}, \citenamefont {Bamberg}, \citenamefont {Ma},
  \citenamefont {Liu}, \citenamefont {He}, \citenamefont {Yan},\ and\
  \citenamefont {Ruhl}}]{Liu2016}%
  \BibitemOpen
  \bibfield  {author} {\bibinfo {author} {\bibfnamefont {B.}~\bibnamefont
  {Liu}}, \bibinfo {author} {\bibfnamefont {J.}~\bibnamefont {Meyer-Ter-Vehn}},
  \bibinfo {author} {\bibfnamefont {K.~U.}\ \bibnamefont {Bamberg}}, \bibinfo
  {author} {\bibfnamefont {W.~J.}\ \bibnamefont {Ma}}, \bibinfo {author}
  {\bibfnamefont {J.}~\bibnamefont {Liu}}, \bibinfo {author} {\bibfnamefont
  {X.~T.}\ \bibnamefont {He}}, \bibinfo {author} {\bibfnamefont {X.~Q.}\
  \bibnamefont {Yan}}, \ and\ \bibinfo {author} {\bibfnamefont
  {H.}~\bibnamefont {Ruhl}},\ }\href {\doibase
  10.1103/PhysRevAccelBeams.19.073401} {\bibfield  {journal} {\bibinfo
  {journal} {Physical Review Accelerators and Beams}\ }\textbf {\bibinfo
  {volume} {19}},\ \bibinfo {pages} {1} (\bibinfo {year} {2016})}\BibitemShut
  {NoStop}%
\bibitem [{\citenamefont {Liu}\ \emph {et~al.}(2018)\citenamefont {Liu},
  \citenamefont {Meyer-Ter-Vehn},\ and\ \citenamefont {Ruhl}}]{Liu2018}%
  \BibitemOpen
  \bibfield  {author} {\bibinfo {author} {\bibfnamefont {B.}~\bibnamefont
  {Liu}}, \bibinfo {author} {\bibfnamefont {J.}~\bibnamefont {Meyer-Ter-Vehn}},
  \ and\ \bibinfo {author} {\bibfnamefont {H.}~\bibnamefont {Ruhl}},\ }\href
  {\doibase 10.1063/1.5051317} {\bibfield  {journal} {\bibinfo  {journal}
  {Physics of Plasmas}\ }\textbf {\bibinfo {volume} {25}} (\bibinfo {year}
  {2018}),\ 10.1063/1.5051317}\BibitemShut {NoStop}%
\bibitem [{Note1()}]{Note1}%
  \BibitemOpen
  \bibinfo {note} {This shift of the peak is (at least partially) explained by
  two major differences between the simulated system in 1D and 3D, that affect
  the {RTF-RPA} process, though it is inherently a 1D phenomenon. With added
  transversal dimensions, the relativistic density can be reduced not just by
  electron relativistic mass increase, but also number density decrease due to
  the transverse ponderomotive force; also, self focusing occurs that can
  increase $a_0$ quite significantly, which in turn also shifts the peak to
  higher bulk densities.}\BibitemShut {Stop}%
\bibitem [{Note2()}]{Note2}%
  \BibitemOpen
  \bibinfo {note} {Naively, this an intuitive explanation: The plasma in the
  downstream region is heated due to accelerated electrons. Far from the {RTF}
  this effect vanishes and $n_0/(T_{e,\infty }+1)\approx n_0$, while directly
  at the front $n_0/(T_{e,0}+1)\approx 0$. The laser has to push against this
  downstream gradient, i.e. has to work against the average $n_0/2$, hence
  $P\equiv 2/(T_{e}+1)\approx 1$. The mathematical derivation can be found in
  the appendix.}\BibitemShut {Stop}%
\bibitem [{\citenamefont {Bussmann}\ \emph {et~al.}(2013)\citenamefont
  {Bussmann}, \citenamefont {Burau}, \citenamefont {Cowan}, \citenamefont
  {Debus}, \citenamefont {Huebl}, \citenamefont {Juckeland}, \citenamefont
  {Kluge}, \citenamefont {Nagel}, \citenamefont {Pausch}, \citenamefont
  {Schmitt}, \citenamefont {Schramm}, \citenamefont {Schuchart},\ and\
  \citenamefont {Widera}}]{Bussmann2013}%
  \BibitemOpen
  \bibfield  {author} {\bibinfo {author} {\bibfnamefont {M.}~\bibnamefont
  {Bussmann}}, \bibinfo {author} {\bibfnamefont {H.}~\bibnamefont {Burau}},
  \bibinfo {author} {\bibfnamefont {T.~E.}\ \bibnamefont {Cowan}}, \bibinfo
  {author} {\bibfnamefont {A.}~\bibnamefont {Debus}}, \bibinfo {author}
  {\bibfnamefont {A.}~\bibnamefont {Huebl}}, \bibinfo {author} {\bibfnamefont
  {G.}~\bibnamefont {Juckeland}}, \bibinfo {author} {\bibfnamefont
  {T.}~\bibnamefont {Kluge}}, \bibinfo {author} {\bibfnamefont {W.~E.}\
  \bibnamefont {Nagel}}, \bibinfo {author} {\bibfnamefont {R.}~\bibnamefont
  {Pausch}}, \bibinfo {author} {\bibfnamefont {F.}~\bibnamefont {Schmitt}},
  \bibinfo {author} {\bibfnamefont {U.}~\bibnamefont {Schramm}}, \bibinfo
  {author} {\bibfnamefont {J.}~\bibnamefont {Schuchart}}, \ and\ \bibinfo
  {author} {\bibfnamefont {R.}~\bibnamefont {Widera}},\ }in\ \href {\doibase
  10.1145/2503210.2504564} {\emph {\bibinfo {booktitle} {International
  Conference for High Performance Computing, Networking, Storage and Analysis,
  SC}}},\ \bibinfo {series and number} {SC '13},\ \bibinfo {organization}
  {ACM}\ (\bibinfo  {publisher} {ACM},\ \bibinfo {address} {New York, NY,
  USA},\ \bibinfo {year} {2013})\ pp.\ \bibinfo {pages} {1--12}\BibitemShut
  {NoStop}%
\end{thebibliography}
\end{document}